\documentclass{emulateapj}

\newcommand{\citepeg}[1]{\citep[{e.g.,}][]{#1}}
\def\Swift{{\textit{Swift}}\,}

\shorttitle{SHOALS II: NIR Luminosity Distribution}
\shortauthors{Perley et al.}

\begin{document}

\title{The Swift GRB Host Galaxy Legacy Survey--- \\II. Rest-Frame NIR Luminosity Distribution and Evidence for a Near-Solar \\ Metallicity Threshold}

\def\cit{1}
\def\mail{*}
\def\dark{2}
\def\leicester{3}
\def\harvard{4}
\def\ipac{5}
\def\iaa{6}
\def\eso{7}
\def\warwick{8}
\def\edinburgh{9}
\def\puc{10}
\def\mia{11}

\author{D.~A.~Perley\altaffilmark{\cit,\dark,\mail}, 
N.~R.~Tanvir\altaffilmark{\leicester},
J.~Hjorth\altaffilmark{\dark}, 
T.~Laskar\altaffilmark{\harvard},
E.~Berger\altaffilmark{\harvard}, \\
R.~Chary\altaffilmark{\ipac},
A.~de~Ugarte~Postigo\altaffilmark{\iaa},
J.~P.~U.~Fynbo\altaffilmark{\dark},
T.~Kr\"uhler\altaffilmark{\dark,\eso}, 
A.~J.~Levan\altaffilmark{\warwick}, \\
M.~J.~Micha{\l}owski\altaffilmark{\edinburgh}, and
S.~Schulze\altaffilmark{\puc,\mia}
}

\altaffiltext{\cit}{Department of Astronomy, California Institute of Technology, MC 249-17, 1200 East California Blvd, Pasadena CA 91125, USA}
\altaffiltext{\dark}{Dark Cosmology Centre, Niels Bohr Institute, University of Copenhagen, Juliane Maries Vej 30, 2100 K{\o}benhavn {\O}, Denmark}
\altaffiltext{\leicester}{Department of Physics and Astronomy, University of Leicester, University Road, Leicester, LE1 7RH, UK}
\altaffiltext{\harvard}{Harvard-Smithsonian Center for Astrophysics, 60 Garden Street, Cambridge, MA 02138, USA}
\altaffiltext{\ipac}{US Planck Data Center, MS220-6, Pasadena, CA 91125, USA}
\altaffiltext{\iaa}{Instituto de Astrof\'isica de Andaluc\'ia (IAA-CSIC), Glorieta de la Astronom\'ia s/n, E-18008, Granada, Spain}
\altaffiltext{\eso}{European Southern Observatory, Alonso de C\'ordova 3107, Vitacura, Casilla 19001, Santiago 19, Chile}
\altaffiltext{\warwick}{Department of Physics, University of Warwick, Coventry CV4 7AL, UK}
\altaffiltext{\edinburgh}{Scottish Universities Physics Alliance, Institute for Astronomy, University of Edinburgh, Royal Observatory, Edinburgh, EH9 3HJ, UK}
\altaffiltext{\puc}{Instituto de Astrof\'isica, Facultad de F\'isica, Pontificia Universidad Cat\'olica de Chile, Vicu\~{n}a Mackenna 4860, 7820436 Macul, Santiago 22, Chile}
\altaffiltext{\mia}{Millennium Institute of Astrophysics, Vicu\~na Mackenna 4860, 7820436 Macul, Santiago, Chile}
\altaffiltext{\mail}{e-mail: dperley@dark-cosmology.dk}

\slugcomment{Accepted to ApJ 2015-10-31}

\begin{abstract}
We present rest-frame NIR luminosities and stellar masses for a large and uniformly-selected population of GRB host galaxies using deep Spitzer Space Telescope imaging of 119 targets from the Swift GRB Host Galaxy Legacy Survey spanning $0.03 < z < 6.3$, and determine the effects of galaxy evolution and chemical enrichment on the mass distribution of the GRB host population across cosmic history.  We find a rapid increase in the characteristic NIR host luminosity between $z\sim0.5$ and $z\sim1.5$, but little variation between $z\sim1.5$ and $z\sim5$.    Dust-obscured GRBs dominate the massive host population but are only rarely seen associated with low-mass hosts, indicating that massive star-forming galaxies are universally and (to some extent) homogeneously dusty at high-redshift while low-mass star-forming galaxies retain little dust in their ISM.  Comparing our luminosity distributions to field surveys and measurements of the high-$z$ mass-metallicity relation, our results have good consistency with a model in which the GRB rate per unit star-formation is constant in galaxies with gas-phase metallicity below approximately the Solar value but heavily suppressed in more metal-rich environments.  This model also naturally explains the previously-reported ``excess'' in the GRB rate beyond $z\gtrsim2$; metals stifle GRB production in most galaxies at $z<1.5$ but have only minor impact at higher redshifts.  The metallicity threshold we infer is much higher than predicted by single-star models and favors a binary progenitor.   Our observations also constrain the fraction of cosmic star-formation in low-mass galaxies undetectable to Spitzer to be small at $z<4$. 
\end{abstract}

% S1
\section{Introduction}
\label{sec:intro}

Long-duration gamma-ray bursts provide a unique and powerful means of studying galaxy evolution: as the extremely luminous explosions of massive stars \citep{WoosleyBloom2006}, they are detectable out to very high redshifts \citep{Tanvir+2009,Salvaterra+2009,Cucchiara+2011}, provide backlights with which to study the intervening ISM and IGM in detail \citepeg{Prochaska+2007,Prochaska+2009,Fynbo+2009} and---perhaps most importantly---their rate is governed (at least in part) by that of star-formation.  The redshift distribution of GRBs and the properties of their hosts can therefore be used to study the evolution of the cosmic star-formation rate density with time \citepeg{Totani1997,Wijers+1998,Blain+2000,Porciani+2001} and characterize the sites where stars were formed in the early universe, independent of many of the foils that hinder traditional techniques such as uncertain dust corrections or incompleteness to low-luminosity galaxies \citepeg{Djorgovski+2001,RamirezRuiz+2002,Tanvir+2012,Trenti+2012,Schulze+2015,Greiner+2015}.

However, only a tiny minority of massive stars actually produce a gamma-ray burst, and as a result the GRB rate may in principle be influenced significantly by factors other than purely the rate of star-formation, such as metallicity.  A detailed understanding of any such variations is essential to apply GRB-based inferences of the cosmic star-formation rate density or other broader topics with confidence.  Furthermore, they would serve to illuminate the nature of the GRB progenitor: the simplest single-star progenitor formation scenarios imply a strong preference or even requirement for very low metallicities, due to the need to avoid loss of mass and angular momentum in line-driven winds \citepeg{Heger+2000,Hirschi+2005,Yoon+2005,Yoon+2006,Woosley+2006}.  Binary-evolution models are more flexible and typically imply relatively modest metal sensitivity \citepeg{Izzard+2004,Fryer+2005,Podsiadlowski+2010}.

The extent to which the GRB rate varies as a function of environmental factors such as metallicity remains an open question observationally. Studies of the redshift distribution alone provide some insight:  the comoving GRB rate density exhibits qualitatively similar behavior as the star-formation history over most of cosmic time, but shows a modest surplus of GRBs at high-redshift ($z>2-3$) relative to what would be expected if the GRB rate tracked the star-formation rate exactly \citepeg{Kistler+2008,Butler+2010,Wanderman+2010,Robertson+2012,Jakobsson+2012,Lien+2014,Paper1}.  This alone seems to rule out \emph{extremely} metal-dependent production models and provides some evidence in favor of a more modest dependence.
However, this technique is necessarily imprecise: the comoving GRB rate measured at any redshift is an average over all galaxies at that epoch (a diverse population spanning orders of magnitude in metallicities, star-formation rates, etc.) and cannot reveal which types of galaxies contribute most (or least) to the GRB rate.  A thorough understanding of the link between GRB production and environment requires also characterizing of the population of GRB host galaxies directly.  

Many host-galaxy studies have been performed over the past decade; a summary of these efforts can be found in \citealt{Paper1} (hereafter, Paper I) and in the review of \cite{Levesque+2014}.  Nearby ($z<1$) GRBs are commonly found in hosts that are blue, young, low-mass and metal-poor in comparison to the star-forming field population, but not necessarily extreme in regards to any of these attributes \citep{Savaglio+2009}.  They occur only very rarely in metal-rich and massive galaxies in this redshift range: large, massive spirals with super-Solar metallicity similar to the Milky Way contribute substantially to the local star-formation rate density but are almost never observed to host GRBs \citepeg{LeFloch+2006,Graham+2013,Vergani+2015}.

The situation at higher redshifts is much less clear.  If the rate of GRBs is controlled by metallicity, the declining metallicities of galaxies with increasing redshift should result in a high-redshift host population that is much more representative of star-formation overall compared to $z\sim$ 0 \citep{Fynbo+2006a,Kocevski+2009}.  The first deep NIR and \emph{Spitzer} studies of GRB hosts at $z>1$ produced few detections at these wavelengths, suggesting a host population that remained quite low in average mass \citep{LeFloch+2003,LeFloch+2006,CastroCeron+2010,Laskar+2011}.  However, these early samples generally under-represented or omitted dust-obscured bursts: more recently, significant numbers of GRBs in massive and metal-rich galaxies have been found by targeting dust-obscured ``dark'' bursts specifically \citep{Kruehler+2012,Rossi+2012,Perley+2013a,Hunt+2014}.  Also, even without specifically including dark bursts, the UV luminosity distribution of small samples of GRB hosts at $z=2-4$ seems to show reasonable consistency with a population that selects galaxies in proportion to star-formation (\citealt{Fynbo+2008,Chen+2009,Greiner+2015,Schulze+2015}).

Actually extending these results to firm conclusions about the properties of the high-$z$ host population or the nature of the progenitor is challenging: high-redshift massive GRB hosts clearly exist, but their actual abundance, and where they fit in the overall distribution of properties of host galaxies at $z>1$, is difficult to quantify, and the topic is still debated \citep{Savaglio+2012,Perley+2013a,Hunt+2014}.  Observational studies conducted so far have been heterogeneously-targeted (biased for or against obscured bursts, and sometimes biased in favor of bright and luminous hosts that are easiest to identify and study), limited in depth or wavelength coverage (based on observations in only one or two filters and difficult to connect with real, physical host properties), and/or too small or too limited in redshift to characterize the high-redshift population.

In Paper I we introduced a new survey of the GRB host galaxy population (the Swift Host Galaxy Legacy Survey, or ``SHOALS''), a new project designed to move beyond these limitations.   Our survey provides the first host galaxy sample that is unbiased (homogeneously targeted) and sufficiently large (119 objects) to statistically examine redshift evolution in the host population in detail, and we are actively observing the sample at many different wavelengths and to sufficient depths to detect and thoroughly characterize the properties ($M_*$, SFR, etc.) of each individual host.

In this paper we present our Spitzer 3.6\,$\mu$m photometry of the SHOALS sample.  Spitzer observations are particularly key to the effort, since the rest-frame NIR luminosity probed by Spitzer directly probes the stellar mass---which in turn is tied \citepeg{Tremonti+2004} to the host metallicity.  Metallicity has, for both theoretical and observational reasons, traditionally been viewed as likely to be the primary driver controlling the GRB rate as a fraction of the rate of star formation (the GRB efficiency), although it is not the only candidate.

Our Spitzer-IRAC observations and data analysis are described in \S \ref{sec:observations}, supported by the redshift measurements and higher-resolution ground and space-based imaging described in Paper I.  We outline our results in \S \ref{sec:results}, showing the near-IR luminosity distribution (a good proxy for the mass distribution) of GRB hosts and its evolution with redshift from the local Universe out to redshift 6.  We discuss the connection between host properties and the degree of obscuration of the afterglow, and its implications for future surveys as well as the nature of the ISM in high-redshift galaxies.  We attempt to model the redshift evolution of the host population using a simple luminosity-dependent cutoff model and show that the near-IR properties and redshift distribution of the population are well-described by a simple model in which GRBs are heavily suppressed above a maximum metallicity (oxygen abundance) threshold of 12 + log[O/H] = 8.94 $\pm$ 0.04 using the scale of \citealt{Kobulnicky+2004} (KK04).  We compare our results to previous work and discuss implications in \S \ref{sec:conclusions}.

% S2
\section{Observations and Analysis}
\label{sec:observations}

% Figure 1
\begin{figure*}
\centerline{
\includegraphics[scale=0.38,angle=0]{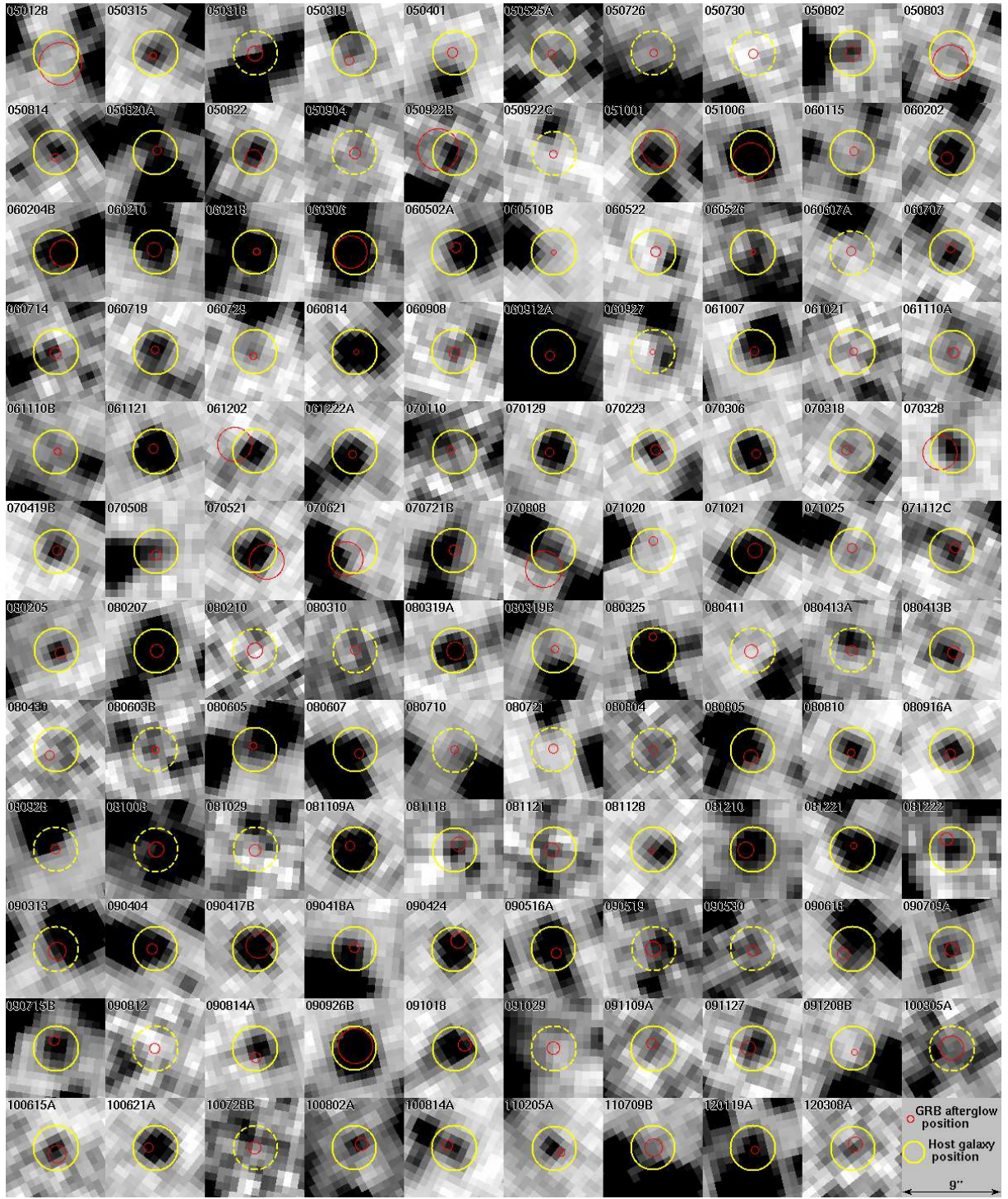}} 
\caption{Spitzer-IRAC 3.6$\mu$m imaging of all 119 uniformly-selected GRB host galaxies from the SHOALS sample.  The red circle denotes the best-available position of the GRB afterglow from the sources identified in Table 3 in Paper I; the yellow position is centered on the GRB host galaxy (Table \ref{tab:spitzerphot}; the center location is measured from optical imaging where possible, otherwise from the Spitzer data itself) or on the afterglow location if no host galaxy is detected at any waveband (the latter case is represented by a dashed aperture).  Due to Spitzer's depth and large PSF, many of the host galaxies are moderately to severely blended with neighboring objects.  Images are 9$^{\prime\prime}$ across; North is up and East is left.}
\label{fig:mosaic}
\end{figure*}

% Figure 2
\begin{figure*}
\centerline{
\includegraphics[scale=0.38,angle=0]{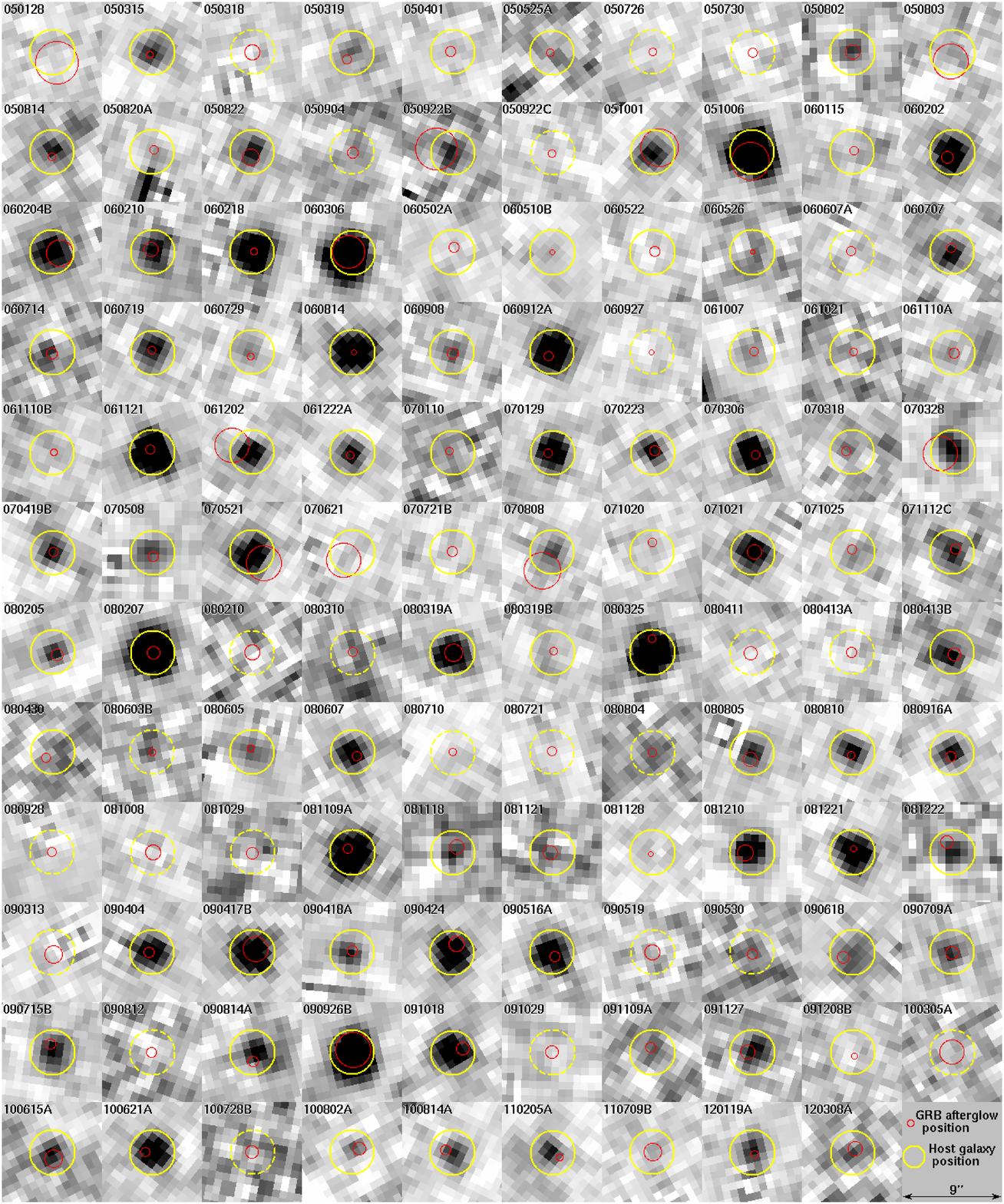}} 
\caption{Same as Figure \ref{fig:mosaic}, but after subtraction of nearby sources using our {\tt galfit}-based scripting procedure to isolate the host galaxy.}
\label{fig:submosaic}
\end{figure*}

% S2.1
\subsection{IRAC Observations}

The Spitzer Infrared Array Camera (IRAC, \citealt{Fazio+2004}) is a four-channel, mid-infrared imager operating on the Spitzer Space Telescope \citep{Werner+2004}.  Since the end of the cryogenic mission in 2009, only two of the channels are functional: channel 1, operating between 3.2--3.9\,$\mu$m; and channel 2, operating between 4.0--5.0\,$\mu$m.  (In this paper we will refer interchangeably to the channels using their channel numbers and central wavelengths:   3.6\,$\mu$m for channel 1 and 4.5\,$\mu$m for channel 2.)

Prior to the start of our project, we examined the Spitzer observations catalog to determine which events within the SHOALS uniform sample (outlined in detail in Paper I) had previously been observed by the telescope.  A large number (almost half) had observations already present in the archive (see Table \ref{tab:spitzerphot} for details).   We did not request re-observations of these targets. For the remaining targets we conducted new observations as part of our Cycle 9 Large Program (GO-90062; PI Perley).  

Our observational strategy was based on the redshift as measured at the time of the start of our campaign (Fall 2012), with deeper observations used for more distant targets.  Integration times were typically 0.2--0.5 hours for $z<1$; 1.5 hours for $1<z<2$; 3 hours for $2<z<3$, and 6 hours for $z>3$.  For events at unknown redshift at the time of the proposal, we integrated for 1.5 hours.  In each case a 100~s frame time was used, dithering using a medium-step dither pattern except in a few cases where the desire to avoid a nearby bright star favored a more compact dithering arrangement.
For targets at unknown redshifts and for those thought to be dark/dust-obscured, we also obtained observations in channel 2 (4.5\,$\mu$m) using the same exposure prescription as for the channel 1 observations.

We downloaded the PBCD (Post-Basic Calibrated Data) observations as they became available in the Spitzer Legacy Archive.  We use the default astrometry provided by the pipeline without further additional alignment, effectively establishing the IRAC images (which are aligned against 2MASS with an accuracy of 0.3$^{\prime\prime}$ by the IPAC pipeline) as the reference system for the survey.

% Figure 3
\begin{figure*}
\centerline{
\includegraphics[scale=0.8,angle=0]{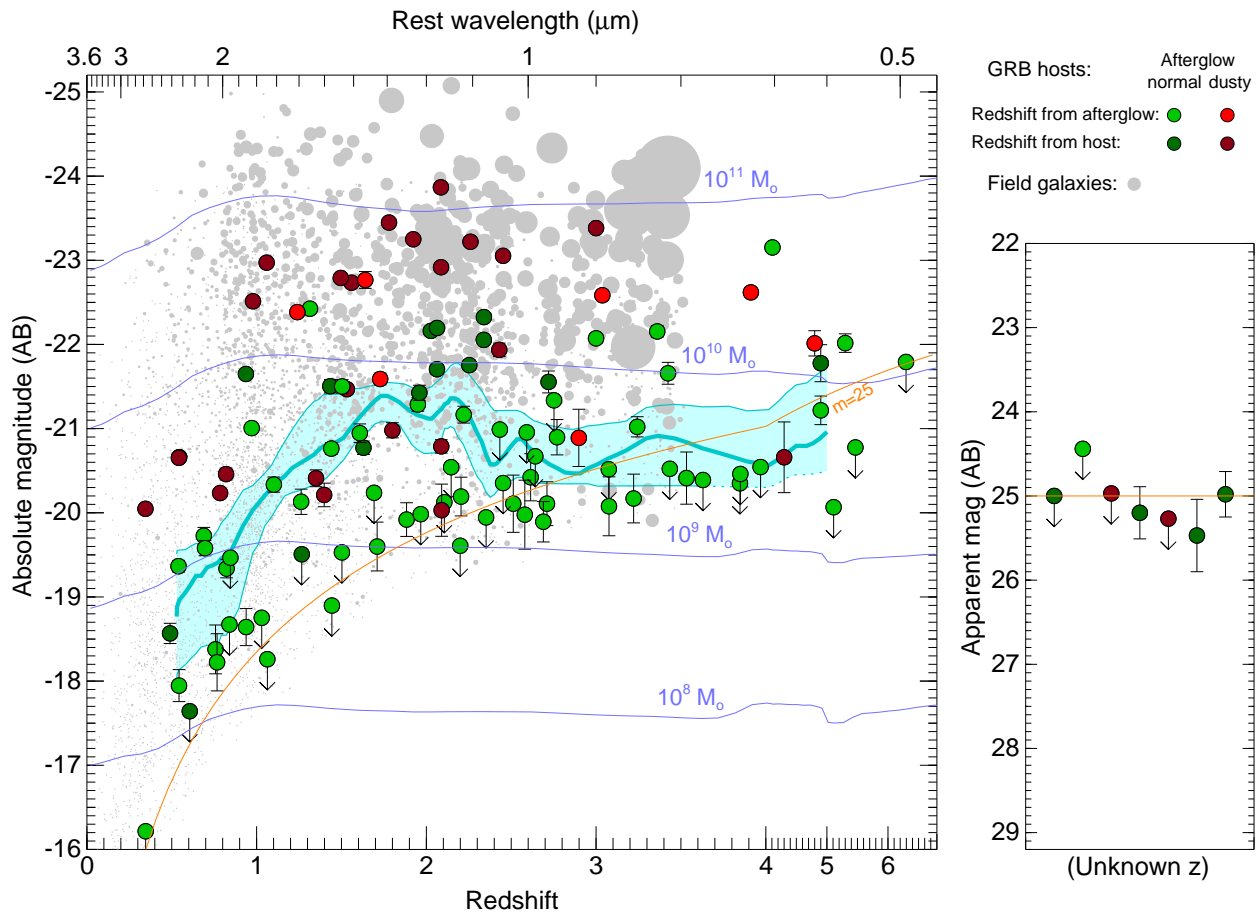}} 
\caption{Near-infrared absolute magnitudes of a uniform sample of 119 GRB host galaxies (110, or 93\%, with measured redshift) and field galaxies.  Magnitudes are AB and correspond to a rest-frame wavelength of $\lambda_{\rm rest}$ = 3.6\,$\mu$m/(1+$z$), the rest-frame equivalent of the IRAC channel 1 filter.  Green points indicate bursts not known to be obscured; red points indicate dust-obscured and ``dark'' bursts.  Darker shades of both colors indicate redshifts measured from late-time host galaxy observations, while redshifts of lighter-shaded points are measured from target-of-opportunity afterglow observations.  $K_s$-selected field galaxies from \cite{Kajisawa+2011} are plotted in gray with area scaled according to SFR.  The right panel shows GRBs at unknown redshift (arbitrary x-axis); in this case colors indicate the source of the redshift upper limit.  Most of these limits are close to $m \sim 25$, which is translated to the main panel as an orange curve.  GRB\,070808 is shown at $z=1.35$.  "The cyan curve shows the redshift-dependent median magnitude of the GRB host sample, with the shaded region denoting the 1$\sigma$ bootstrap uncertainty on this value.}
\label{fig:absmagz}
\end{figure*}

% S2.2
\subsection{Deblending}
\label{sec:subtraction}

Source confusion and contamination are common in even moderately deep Spitzer imaging due to the instrument's large PSF ($\sim1.8$\arcsec).  Indeed, inspection of our images (Figure \ref{fig:mosaic}) demonstrates that a significant fraction of the target host galaxies are contaminated to some degree by flux from neighboring, unrelated objects.  Accurate photometry requires removing this contamination.

We manually examined each Spitzer image simultaneously with any higher-resolution ground-based or HST imaging available of the field (see Paper I) to identify the host-galaxy target and any nearby objects which may contaminate the host or sky apertures.  Using a custom DS9/python script, we drew a fitting box around each host position and estimated the locations and sizes of all detected sources inside it (as well as those of any bright sources outside the box edges that contribute significant flux within the box).  These were then sent as initial inputs via a python wrapper to \texttt{galfit} \citep{Peng+2002} to iteratively fit and subtract all sources within the box.  The default PBCD image is used as the fitter input image with the PBCD uncertainty map providing the pixel uncertainties.   For the input PSF we use our own merged combination of the core and extended PRF files from the IRAC website\footnote{http://irsa.ipac.caltech.edu/data/SPITZER/docs/irac/calibrationfiles/psfprf/}, with the final PSF oversampled by a factor of 5 relative to the PBCD image resolution (or by a factor of 10 relative to the native pixel scale.)   We use a Sersic model with index fixed at $n=1$ (typically circular, but elliptical if the source is visibly elongated) or a point-source model depending on the appearance and brightness of the target.   The residual images were inspected after the fit, and the routine was re-run with new inputs if necessary until all sources near the host galaxy (including the host itself, if well-detected) were removed as completely as possible from the final residual map.  We then created a final output image by constructing a {\tt galfit} model using the best-fit parameters for all components \emph{except} for those associated with fitting the host itself, and subtracted this model from the initial PBCD image to isolate the host galaxy.  The subtracted maps are shown in Figure \ref{fig:submosaic}.  This procedure was effective at removing contaminating flux for all of our fields, usually leaving negligible residuals except in the cores of very bright stars and galaxies.

% S2.3
\subsection{Host-Galaxy Photometry}
\label{sec:photometry}

Our \texttt{galfit} procedure returns calibrated PSF-based magnitudes using the appropriate zeropoints for each image.   For our final photometry, however, we employ the procedure recommended in the IRAC handbook of calculating aperture photometric magnitudes.  These are calculated via the subtracted images described above using a custom IDL wrapper around the \texttt{aper} procedure in the Astronomy User's Library, employing an aperture radius of 1.8\arcsec\ (1.5 native pixels or 3 resampled pixels).  Zeropoints (including aperture corrections) are based on the corresponding values reported in the Spitzer-IRAC handbook, interpolated using our PSF model (\S\ref{sec:subtraction}) to permit the use of fractional native-pixel aperture sizes not given in the handbook.  Apertures are centered on the host-galaxy localization as determined from our ground-based imaging or from the fitting/subtraction procedure; the aperture centers employed are given in Table \ref{tab:spitzerphot}.  We estimate uncertainties via the scatter of pixel values in the sky annulus (within the \texttt{aper} procedure), rescaled using an empirical correction for correlated noise expected in subpixel-sampled images.   We measured this correction to be a factor of 1.25 (in flux) for our PBCD images by placing a large number of random apertures in blank regions of several non-confusion-limited images and calculating the true scatter in the resulting flux values.  Error estimates do not explicitly include systematics associated with the subtraction procedure, which are expected to be minor.

In many cases the host galaxy is not detected in the IRAC imaging.  For these cases, we calculate an upper limit.  If the host is detected in ground-based imaging we fix the aperture location at the host position and calculate a 2$\sigma$ upper limit on the flux at that location.   If we do not know the host location, we center the aperture on the best-available afterglow position and calculate a $3\sigma$ upper limit.  (The higher $\sigma$ threshold reflects the additional uncertainty associated with not knowing exactly where to place the aperture.)

Photometry on some of the hosts presented here were also published in previous work (in particular \citealt{Laskar+2011} and \citealt{Perley+2013a}) but in all cases the analysis here is independent, using the semi-automated subtraction and photometry procedure above consistently for all targets.  Our photometry is generally consistent with these previously published values.  We note that further observations and analysis (better de-confusion and precise localization of additional host locations using deeper images) may provide additional improvements in the future so the values reported here do not necessarily represent the final values for the survey, but we do not expect large changes and the current photometry should be adequate for the purposes of this paper.

% old S2.4 - no longer necessary

% Figure 4
\begin{figure}
\centerline{
\includegraphics[scale=0.65,angle=0]{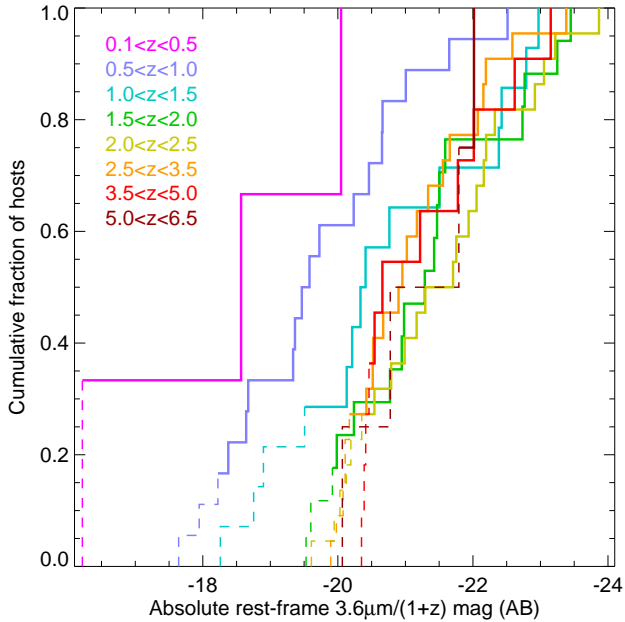}} 
\caption{Cumulative luminosity distribution of GRB hosts in a variety of redshift bins spanning most of cosmic history.  Dashed lines indicate upper limits (upper limits are plotted as 1 magnitude fainter than the measured limiting value; this portion of the curve should not be taken as representative of the true distribution.)  The host galaxy luminosity function increases sharply from $z\sim0.25$ to $z\sim1.5$ and then changes relatively little out to higher redshifts.}
\label{fig:lumdistz}
\end{figure}

% S2.4 (previously 2.5)
\subsection{Mass Conversion}
\label{sec:massconv}

The luminosity of a galaxy as measured in the IRAC 3.6$\mu$m band is a good tracer of stellar mass even out to high redshifts, since it probes emission redward of the Balmer break (dominated by evolved stars and only moderately dependent on population age) at nearly all redshifts.  While a precise stellar mass estimate requires an SED fit to many photometric data points (to incorporate age, extinction, and similar dependencies), a reasonable estimate can be obtained from Spitzer observations alone.

For each of our (known-redshift) galaxies, we calculate the the quantity $M_{\rm 3.6/(1+z)}$ = $m_{\rm obs, 3.6} - \mu(z) + 2.5\rm{log}_{10}(1+z)$; this is the luminosity (AB absolute magnitude) of the host galaxy at the \emph{rest} wavelength observed by the 3.6\,$\mu$m filter after cosmological redshift, i.e., the absolute magnitude at a wavelength of $3.6/(1+z)$ $\mu$m.   This quantity alone can be used as a reasonable stellar-mass proxy, but we also calculate stellar mass estimates directly using the following prescription.  For each decade in stellar mass ($10^7 M_\odot$, $10^8 M_\odot$, etc.) and for a grid of different redshifts between $z=0-10$, we create a model galaxy SED by summing \cite{bc03} galaxy SED templates for a continuous star-formation history between the present time and a maximum age determined by the mass (100 Myr for $\leq 10^8 M_\odot$, 300 Myr for $10^9 M_\odot$, 700 Myr for $10^{10} M_\odot$, and 3 Gyr for $\geq 10^{11} M_\odot$; if this age is greater than the age of the Universe at the template redshift we instead use the age of the Universe as the maximum age).  We assume modest dust attenuation ($A_V$ of 0.05 mag for $\leq 10^8 M_\odot$, 0.1 mag for $10^{9-10} M_\odot$, and 0.3 mag for $\geq 10^{11} M_\odot$)\footnote{These preferred values of $A_V$ and age were chosen by comparing the resulting $M_{\rm 3.6/(1+z)}(M*, z)$ conversion function with the SED-fit masses and 3.6$\mu$m magnitudes in the MODS and Ultra-VISTA surveys; i.e., our SED model was developed to ensure that we could accurately predict the average masses of MODS/Ultra-VISTA galaxies from just their 3.6$\mu$m magnitudes.  Indeed, single-band masses via our procedure for individual galaxies using IRAC photometry from the comparison field surveys show good agreement with masses derived from a full SED fit, with an additional scatter of a factor of 2--3 (equivalent to 0.3--0.5 dex).} and a \cite{Chabrier+2003} IMF.
For each SED we then calculate via synthetic photometry the same quantity which we measure using Spitzer, $M_{\rm 3.6/(1+z)}$.  The observed magnitudes can then be translated to stellar masses via interpolation on the mass-redshift grid.

Compared to, e.g., scaling based on the rest-frame $K$-band luminosity, this empirical procedure avoids the use of $k$-corrections and allows for the expected not-fully-linear scaling of luminosity with mass (because lower-mass galaxies tend to be younger and have lower mass-to-light ratios.)  Of course, like all mass derivation procedures, it is subject to various assumptions related to the choice of stellar templates, IMF, star-formation history, etc.   In general, in this work we will deal with the raw observed luminosities (from which we can compare our host galaxies to each other and to observed stellar populations free of any assumptions) at most stages of the analysis, converting to mass only to add physical context or interpret our results.

% S3
\section{Results}
\label{sec:results}

% S3.1
\subsection{Near-Infrared Luminosity Distribution and Redshift Evolution}

In Figure \ref{fig:absmagz} we plot the near-infrared luminosity (see above) of all galaxies in our survey as a function of redshift.  Blue curves show the magnitude that would be measured for several galaxy templates of various masses as a function of redshift using the templates described in the previous section.  
A few GRBs have been treated specially in this plot and the ensuing analysis.  For the initial 3.6\,$\mu$m observation of GRB\,050401 in Cycle 4, the orientation of the detector resulted in a great deal of stray light from a bright nearby star; this target was re-observed in Cycle 8 at a better orientation but at 4.5\,$\mu$m only; because this image is deeper (and the expected 3.6--4.5\,$\mu$m AB color close to 0.0) we use the 4.5\,$\mu$m magnitude.  The region near the afterglow position of GRB\,090313 is strongly blended with a foreground galaxy and is omitted from the plot and subsequent analysis (pending acquiring deep ground-based data to localize the host centroid).   GRB\,080310 is affected by a diffraction spike from a nearby star; we attempted to model and subtract it, but it is difficult to evaluate the reality of sources at or near the host location and a conservative estimate of the field depth does not reach our program goals, so we omit this also.  

The host galaxy luminosity distribution exhibits strong evolution with redshift, at least at the low-$z$ end: luminous galaxies (e.g. those with $M<-21$) are common above $z>1.5$ but effectively vanish from the host sample below $z<1$.  To better quantify this, we calculated the median host magnitude as a function of redshift using a moving window encompassing up to 21 objects (10 at lower redshifts and 10 at higher redshifts).  Nondetections are included; since the median magnitude is always above the limits of typical nondetectons, our lack of knowledge of their actual flux does not affect the median calculation. 
The uncertainty of this running median is calculated (and shown as the filled region on the diagram) via a simple resample-with-replacement bootstrap technique.\footnote{The treatment of nondetections does affect the faint bound of this uncertainty envelope at $z>3.7$, since we detect just above 50\% of hosts at these redshift and many replacement trials produce a median set by the assumed value of the limits.  A maximally conservative treatment would place the nondetections at a flux level of zero; in this situation the lower uncertainty on the median strictly becomes unbounded.  For Figure \ref{fig:absmagz} we randomly assign nondetections at a flux level between the limiting value and zero, but we indicate the lower bound using a dotted line in this region to indicate that this is dependent on the assumption of the distribution of limiting fluxes.}
The resulting curves are shown in Figure \ref{fig:absmagz} in cyan.  A marked increase of 2--3 magnitudes in the median luminosity is evident up to $z \sim 1.4$, at which point the average magnitude levels off, showing no further (significant) variation out to higher redshifts.  Some of this behavior can be interpreted as the result of the shift of the effective rest-frame wavelength of the fixed filter (see the thin blue equal-mass curves); in particular a modest downturn of up to 1 mag is expected below $z<1$ as a result of the stellar bump moving out of the 3.6\,$\mu$m bandpass.  However, star-forming galaxy SEDs are relatively flat (in AB) between rest-frame wavelengths of 0.8--2.2$\mu$m which most of our observations correspond to, and the strong decline in the typical luminosity (by 2--3 mag) at low redshifts is too large to be interpreted as anything other than intrinsic evolution in the population.  Specifically, the characteristic mass of the host population declines by approximately an order of magnitude from $z\sim1.5$ to the present time.

An alternative visualization of the redshift evolution of the luminosity distribution is shown in Figure \ref{fig:lumdistz}, which plots the cumulative luminosity distribution in a series of redshift bins from $0<z<0.5$ to $4.5<z<6$.  Nondetections are represented arbitrarily as the limiting magnitude plus 1.0, and shown as a dotted line.  While the two lowest-redshift bins show a distribution weighted towards low-luminosity hosts, the remaining curves are nearly identical.

% S3.2
\subsection{Dark Bursts and the Role of Extinction}

The nature of the ``dark'' GRB population and its implications for the GRB host galaxy population overall has been a major topic in the study of GRB environments during the past decade.  Systematic follow-up of the host galaxies of these events has shown that the hosts of the dust-obscured population show dramatically higher star-formation rates and stellar masses and redder colors than the hosts of unobscured GRBs \citepeg{Kruehler+2011,Svensson+2012,Perley+2013a,Rossi+2012,Hunt+2014}, with major implications for the nature of the GRB host population (and its connection to star-formation) overall.

Given these past results, it is unsurprising that we observe a strong correlation between the existence of obscuration affecting the afterglow and the properties of the host, as illustrated in Figure \ref{fig:absmagz}.   It is noteworthy that GRBs occurring within the most NIR-luminous quartile of GRB hosts (those at $M_{3.6/(1+z)} < -22.5$ mag, or a mass of about $M_* > 3\times10^{10} M_\odot$) are almost exclusively moderately- to severely obscured.   This result also emphasizes the critical role of choosing a uniformly-selected sample for studies of this type: had our survey considered only those events with optical afterglow-determined redshifts (light-colored red and green points in the main panel of Figure \ref{fig:absmagz}) our results would be entirely different, since nearly all of the luminous galaxies in the sample hosted GRBs whose redshifts required host-galaxy follow-up.

In contrast, GRBs in the least-luminous galaxies are only rarely obscured.  Among GRBs in our sample at known redshift, only a single example is definitively dark/dusty and hosted within a galaxy with a luminosity-derived mass less than $\sim3 \times 10^{9} M_\odot$.  While three others at \emph{unknown} redshift are in galaxies that are very faint and---unless they are at $z\sim5$---likely hosted in similar galaxies, this is still dwarfed by the numbers of dusty GRBs in massive galaxies or by the numbers of unobscured GRBs in low-mass galaxies.  A dust-obscured afterglow in a low-mass host is a very rare situation.

While a detailed analysis of the coupling between the properties of dust seen towards the GRB afterglow ($A_V$, extinction law, etc.) and host galaxy properties will be reserved for future work, we note in passing that the nearly-ubiquitous presence of obscuration in the afterglows of GRBs originating from luminous systems provides further evidence in support of the notion that dust in massive galaxies is fairly homogeneous, or at least contains a large diffuse component.  It also suggests that ongoing star-formation in dense and dusty clouds in \emph{low}-mass galaxies is not a large contributor to cosmic star-formation and more broadly that low-mass galaxies harbor relatively little dust. (See also our discussion of this topic in \citealt{Perley+2013a}, and the discussion in \citealt{Schady+2014}.)  This is in agreement with conclusions of decreasing dust content towards fainter galaxies in deep surveys \citepeg{Bouwens+2009,Bouwens+2014a,Castellano+2012,Finkelstein+2012,Oesch+2013} but extends these results to include optically-thick, heterogeneously-distributed dust (which would not manifest itself in the galaxy colors) and to galaxies with arbitrarily low masses.

We do not see obvious redshift evolution in the tendency for dust-obscured GRBs to inhabit more luminous galaxies: all but one of the dust-obscured bursts at $3 < z < 5$ lies within a host galaxy with a mass above the median.  This suggests that even over this redshift range, dark GRBs cannot be neglected in studies of host demographics.  Several of these GRBs did have an afterglow bright enough to secure an absorption-based redshift measurement, however---which may suggest that dusty GRBs become less so between $z\sim2.5$ and higher redshifts.  However, since dust-obscured GRBs represent only a modest fraction of the total at any redshift, we do not yet have sufficient number statistics to address the issue definitely.

% S3.3
\subsection{Comparison to Field-Selected Star-Forming Galaxies}

The nature of the GRB host population at any redshift is affected both by the true distribution of star-formation within galaxies as well as any preference that the GRB progenitor may have for particular types of environment, and the primary goal of our survey is to search for and characterize these influences.
A detailed examination of this topic will require completion of the multi-band survey and as such will be addressed in future publications---however, the Spitzer observations alone are already quite informative, since the Spitzer magnitude serves as a good tracer of the stellar mass distribution and can be used to produce model-independent comparisons against star-forming galaxies from galaxy field surveys with Spitzer data.

Following an approach similar to our previous study \citep{Perley+2013a}, in Figure \ref{fig:absmagz} we also plot field galaxies from \cite{Kajisawa+2011}, with the area of each point scaled proportional to the star-formation rate of each object.   These galaxies come from a deep $K_s$-band selected catalog (MODS, the MOIRCS Deep Survey) of the GOODS-North field.  This survey is very deep ($K_s > 24.9$ AB mag at 5$\sigma$) and mass-complete to a similar level as is achieved by our Spitzer observations, and contains a large value-added catalog including dust-corrected UV and 24$\mu$m star-formation rate measurements.  It therefore makes an appropriate field-survey comparison sample for our study, although it is susceptible to cosmic variance as a result of the small field footprint (28 arcmin$^2$ for the ``wide'' catalog employed here).  To mitigate against cosmic variance, we also employed the catalogs derived from deep observations of the UDS field from CANDELS \citep{Galametz+2013,Santini+2015} --- approximately 7 times larger in area, although without the long-wavelength star-formation rate measurements.  Since both these surveys have small numbers of targets and may be incomplete at low redshift ($z<0.5$), we also used the much wider (5400 arcmin$^2$) but shallower COSMOS/Ultra-VISTA catalogs \citep{Muzzin+2013} as a comparison population for the (very limited) sample of GRB hosts in that redshift range.

% Figure 5
\begin{figure*}
\centerline{
\includegraphics[scale=0.75,angle=0]{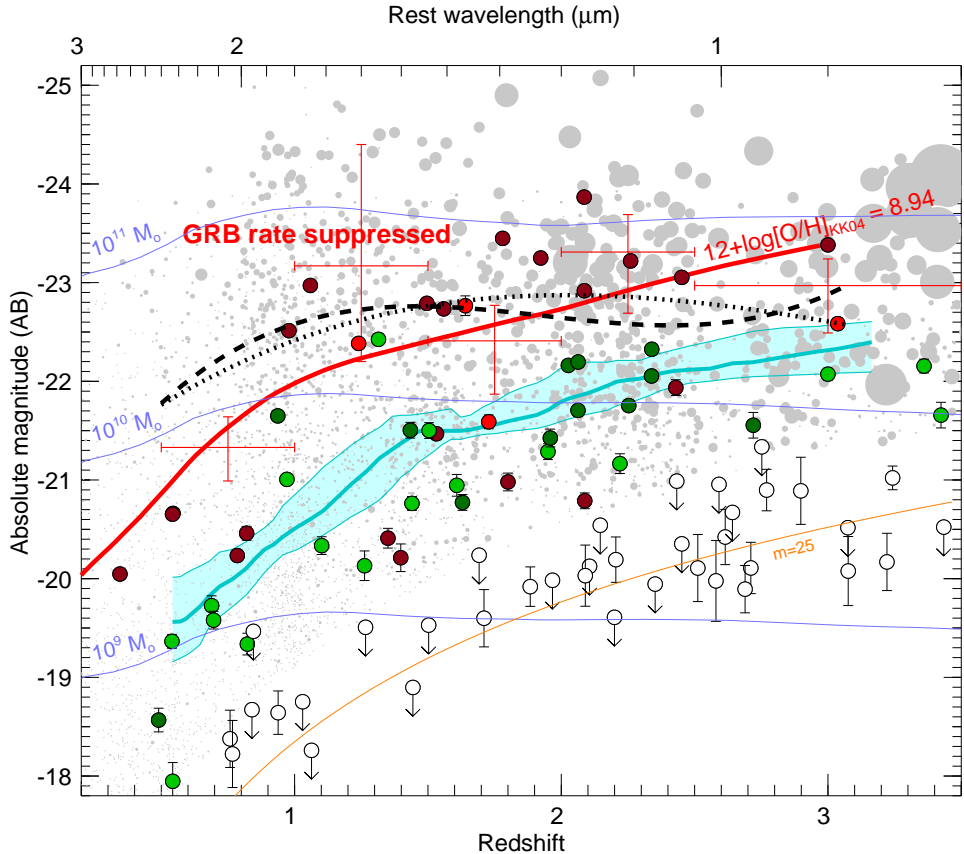}} 
\caption{Same as Figure \ref{fig:absmagz} but now with an apparent-magnitude cut applied to both populations in order to compare the samples only in the regions where both are complete.  The cyan curve again shows the median magnitude of the GRB host population near that redshift (and 1$\sigma$ bootstrap uncertainty), this time only for $m<24.25$ mag detections.  The dashed and dotted curves show the same property calculated from MODS/GOODS-N and CANDELS/UDS, respectively.   GRB hosts are significantly underluminous as a population at all redshifts but especially below $z<1.5$.   The red measurements show our best-fit values for $M_{\rm 3.6/(1+z),thresh}$, a (soft) upper limit for efficient GRB production within a host galaxy (Table \ref{tab:cutvalues}). The solid red curve shows the threshold for a metallicity of log[O/H]=8.94 converted to luminosity (\S \ref{sec:threshold}).}
\label{fig:absmagzcompare}
\end{figure*}

The contribution of undetected galaxies to the cosmic star-formation rate in a field survey is fundamentally uncertain without employing assumptions: extrapolating luminosity functions or relying on theoretical models.  In contrast, GRBs probe star-formation regardless of the detectability of their host galaxies.   To fairly compare the GRB and field galaxy luminosity distributions, we trim our samples at an apparent magnitude cut of $m_{3.6} < 24.25$ AB, which corresponds to the completeness limit of both our host-galaxy survey and of MODS, and examine only the redshift range of $0.5 < z < 3$, since even $K$-band selected samples cannot be complete with respect to rest-frame NIR luminosity at higher redshifts.  (Where relevant we use the aperture-corrected total magnitudes, not the raw catalog magnitudes.)  We then calculate the median absolute magnitude of the host galaxy distribution following the same procedure as in the previous section, and the SFR-weighted median absolute magnitude of the field galaxy sample.  For this purpose, within MODS we use the SFR$_{\rm IR+UV}$ column, which adds UV and 24$\mu$m star-formation rates if detected with Spitzer-MIPS, or otherwise uses an extinction-corrected UV measurement.  For UDS we use the star-formation rate from the ``14a'' SED model (which employs the most flexible star-formation history) and for Ultra-VISTA we employ the SED-fit star-formation rate estimate.  These median curves are plotted as thick solid blue (GRBs) and dashed black lines (MODS) in Figure \ref{fig:absmagzcompare}.  If GRBs were perfect tracers of the cosmic star-formation rate (that is, the GRB rate per unit star-formation was a constant regardless of environmental properties such as metallicity), then the medians of the two distributions would be statistically consistent with each other at every redshift.

% Figure 6
\begin{figure*}
\centerline{
\includegraphics[scale=0.74,angle=0]{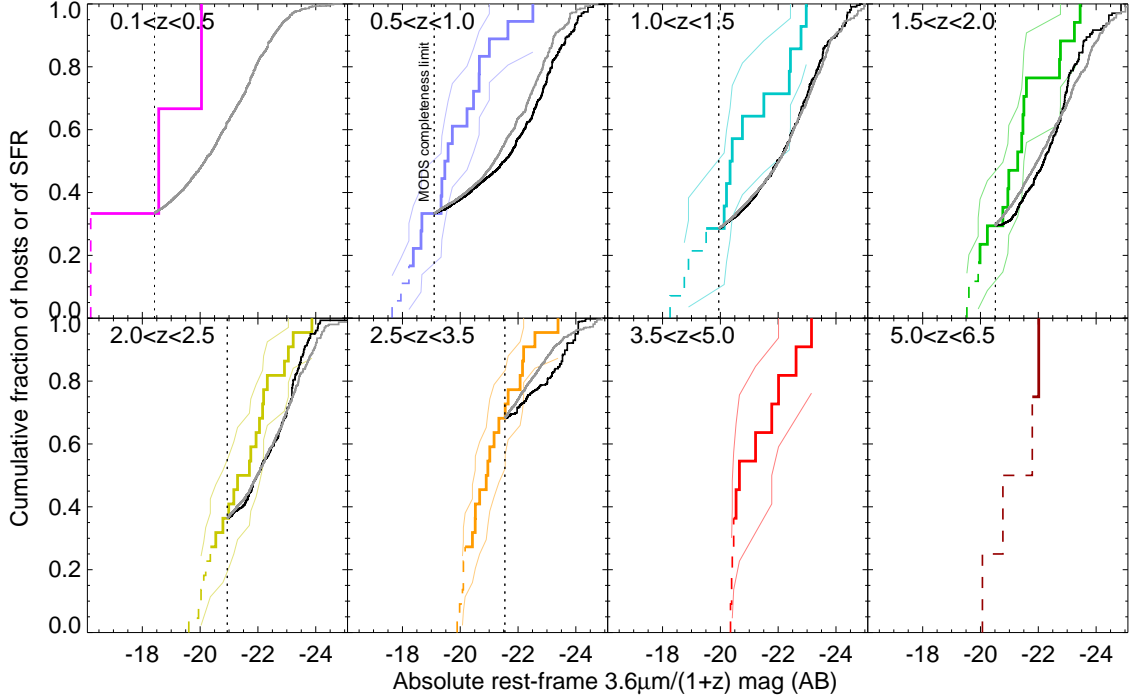}} 
\caption{The cumulative luminosity distribution of GRB hosts (colors) compared to the star-formation-weighted distribution of galaxies in GOODS-N/MODS (black) and UDS/CANDELS (gray) above the completeness limit of the survey.  (For the lowest-redshift bin we instead compare to Ultra-VISTA.)  Since field surveys are incomplete to faint galaxies the galaxy curve is anchored to the host distribution starting at a common limiting magnitude shown as the dotted vertical line.  The curves are highly inconsistent at low redshifts with an almost complete dearth of GRBs in luminous and massive galaxies responsible for 30\% or more of cosmic star-formation at that epoch (as also seen in Figure \ref{fig:absmagz}).  Milder inconsistency is observed at higher redshifts.  No appropriate mass-limited sample exists with which to compare beyond $z>3.5$.  The dashed portion of the GRB curve incorporates nondetections in the same way as Figure \ref{fig:lumdistz}.}
\label{fig:lumdistvssfr}
\end{figure*}

Large differences are observed between the actual average luminosity of our GRB host sample and that of the (SFR-weighted) field galaxy population at low redshift: about 1.5 mag (a factor of four) below $z \sim 1$.  The deviation gradually narrows with increasing redshift but does not completely disappear: at redshifts of $z=2-3$ the GRB host population is still located in galaxies that are, on average, about 0.5 mag fainter than what would be predicted for a strictly uniform tracer.

A more detailed representation of this behavior is shown in Figure \ref{fig:lumdistvssfr}, which provides cumulative distributions for the absolute magnitude of GRB hosts in various redshift ranges, compared to the absolute magnitude distributions \emph{weighted by star formation rate} of galaxies in several field surveys.  The MODS sample is shown as the black line and the UDS sample is shown as a gray line, except in the lowest-redshift bin where we use COSMOS/Ultra-VISTA.  The GRB distribution is heavily weighted towards low-luminosity (low-mass) galaxies out to at least $z\sim1.5$, and skewed more weakly towards higher redshifts.

% S3.4
\subsection{The Threshold Mass and Metallicity for GRB Production}
\label{sec:threshold}

A striking feature of Figures \ref{fig:absmagz} and \ref{fig:lumdistvssfr} is the near-total absence of GRBs in massive systems at low redshift.  Galaxies more massive than $3\times10^{10}$ $M_\odot$ (about half the mass of the Milky Way) are responsible for more than a quarter of the cosmic star-formation rate density at $z\lesssim1$ but produced no GRBs in our sample, which is 100\% complete and unbiased for galaxies in this mass and redshift range.  In contrast, galaxies with a 3.6$\mu$m-derived mass only slightly less ($10^9-10^{10} M_\odot$, within a factor of a few of the LMC) produce the majority of GRBs in this interval.  This suggests two things.  First, it requires that that the GRB rate is a quite sharp function of environment near its critical level: e.g. the GRB efficiency in $10^{10}-10^{11} M_\odot$ galaxies must be roughly an order of magnitude lower than the efficiency within $10^{9}-10^{10} M_\odot$ galaxies.  However, it also indicates that extremely low mass is not necessarily a boon to GRB production: while some extremely faint GRB hosts do exist, the median mass and luminosity are still well within the range that is considered ``normal'' for star-forming galaxies.  Extremely faint galaxies are not required or even particularly favored for GRB prouction.

Our result clearly rules out models in which GRBs are entirely uniform (unbiased) with respect to star-formation, as well as models in which the GRB dependence is relatively smooth.   For example, the model favored by the study of \citealt{Trenti+2015} (largely on the basis of fitting the GRB redshift distribution, but also incorporating some limited host observations) suggests that most ($\sim$85\%) GRBs at $z<1$ form via a metal-independent channel; this would predict a large number of GRBs from massive and luminous galaxies in the nearby universe and is in contradiction to our observations.  (Trenti et al.\ themselves also note that the lack of reported luminous low-$z$ hosts is a point of tension with their model.)

It is, however, in general agreement with previous studies of the GRB host stellar mass distribution at $z<1$.  For example, Figure 3 of \cite{Boissier+2013} also shows an order-of-magnitude drop in GRB efficiency between a host stellar mass of $10^{9.6}$ and $10^{10.4}$ $M_\odot$, while \cite{Vergani+2015} find no GRB hosts at $>10^{10} M_\odot$ in their complete $z<1$ sample (requiring strong supression in this range) but measure a variation in efficiency of only a factor of a few or less below $10^{9.5} M_\odot$ (their Figure 4).  It is also qualitatively consistent with the conclusions of \cite{Svensson+2010}, who compared GRB and cc-SN host properties.

The GRB dependence on host environment is widely speculated to intrinsically be a \emph{metallicity} dependence, in which metal-poor stars form GRBs readily but metal-rich stars much less or not at all.  Due to the complex correlations between different properties in galaxies, it is difficult to be certain from photometric data alone whether or not metallicity (and not some other parameter that also correlates with mass/luminosity) is actually responsible for the mass-dependent GRB rate at low redshift---other possibilities invoke variations in the IMF/binarity associated with starburst intensity, or stellar mergers associated with dynamical interactions found only in very dense/intense star-forming regions \citep{vandenHeuvel+2013}.
Direct spectroscopic metallicities of all of the massive galaxies in our sample would provide a means to evaluate this directly, but we do not (yet) have these observations for our full sample---and even with spectroscopy in hand, ambiguities associated with inhomogeneous conditions in these galaxies can introduce significant complications in interpretation.

However, our sample offers an opportunity to investigate the metallicity-limit hypothesis in a different way.  The mass-metallicity relation evolves strongly with redshift \citepeg{Erb+2006}, leading to a natural prediction \citep{Kocevski+2009} that the threshold stellar mass (and NIR luminosity) should rise with redshift, and do so in a specific way.  Previous studies \citep{Kruehler+2011,Perley+2013a} have hinted that this effect is present but have been unable to make quantitative tests of it due to their incomplete and potentially biased redshift sampling for $z>1$ objects.  With our large, complete sample we are in a position to quantitatively examine this behavior for the first time.

With a simple metallicity-cut model in mind, we use our redshift-subdivided luminosity distributions (Figure \ref{fig:lumdistvssfr}) to test whether the increasing stellar mass with redshift that is apparent in our sample is indeed consistent being the product of a redshift-dependent mass metallicity relation and a redshift-invariant metallicity threshold for GRB production.   For each redshift interval, we resample the field galaxy magnitude distribution such that it is uniform with respect to star-formation (binning individual galaxies of similar magnitudes to produce a new distribution where each magnitude increment has constant total SFR, mirroring the way GRBs are expected to select galaxies) and perform a two-sided K-S test between this and the GRB host luminosity distribution.  This procedure is then repeated for a large number of possible maximum-luminosity thresholds spanning the range of the host sample, recalculating the K-S $p$-value each time.  We then estimate the threshold level, and its upper/lower uncertainties, by marginalization.  These measurements and uncertainties are presented in Table \ref{tab:cutvalues}.

Despite the large size of our survey, we remain strongly limited by the number counts of the GRB host population in each bin and the uncertainty on the threshold measurement is large (typically 0.5--1.0 mag).  However, at all redshift bins at which we can carry out this analysis, a model with no luminosity cut is ruled out ($95-99.99$\% confidence, depending on redshift).  We do, however, find a good match to the distributions at every redshift if a luminosity cut is applied to the field-galaxy sample.

Our measured luminosity thresholds (and uncertainties) can then be translated into stellar masses using the same procedure described in \S \ref{sec:massconv}.  From there, they can be converted into metallicity values using the analytic expressions for the redshift-dependent mass-metallicity relation of \cite{Zahid+2014}.

If the GRB efficiency is controlled only by metallicity (and not other factors that are varying over the Universe's history, such as the average specific SFR) we would expect to measure a consistent metallicity threshold at each redshift.  Indeed, we find that all of our observations are consistent with redshift-independent GRB metallicity threshold of 12 + log[O/H] = 8.94 on the KK04 system with a scatter of only $\sigma = 0.04$ among the different redshift bins.  Converting using the relations of \citealt{Kewley+2008}, this is equivalent to 12 + log[O/H] = 8.64 on the \cite{Pettini+2004} (PP04) system.  In either case this is close to the Solar value ($\sim$8.7, \citealt{Asplund+2009}).

% Table 1
\begin{deluxetable*}{lllllllll}  %*
\tabletypesize{\small}
\tablecaption{Redshift-Dependent Cutoff Values}
\tablecolumns{9}
\tablehead{
\colhead{$z$\tablenotemark{a}} & 
\colhead{$M_{3.6/(1+z)}$\tablenotemark{b}} & 
\colhead{log($\frac{M_*}{M_\odot}$)\tablenotemark{c}} &
\colhead{12+log[O/H]\tablenotemark{d}} &
\colhead{$M_{3.6/(1+z),8.94}$\tablenotemark{e}} & 
\colhead{$f_{\rm SF}$\tablenotemark{f}} &
\colhead{$p_{\rm best}$\tablenotemark{g}} &
\colhead{$p_{\rm 8.94}$\tablenotemark{h}} &
\colhead{$p_{\infty}$\tablenotemark{i}} \\
}
\startdata
0.1--0.5 &                          &                         &                         & -20.30 &      &      &      &       \\
0.5--1.0 & $-21.33^{-0.31}_{+0.34}$ & $ 9.82^{+0.14}_{-0.15}$ & $ 8.91^{+0.04}_{-0.05}$ & -21.53 & 0.33 & 0.71 & 0.43 &$<$0.0001 \\
1.0--1.5 & $-23.17^{-1.23}_{+0.97}$ & $10.70^{+0.57}_{-0.52}$ & $ 9.05^{+0.05}_{-0.11}$ & -22.23 & 0.43 & 0.24 & 0.06 & 0.044 \\
1.5--2.0 & $-22.41^{-0.36}_{+0.54}$ & $10.34^{+0.20}_{-0.30}$ & $ 8.91^{+0.05}_{-0.10}$ & -22.57 & 0.50 & 0.13 & 0.12 & 0.003 \\
2.0--2.5 & $-23.31^{-0.38}_{+0.62}$ & $10.83^{+0.20}_{-0.34}$ & $ 8.99^{+0.04}_{-0.09}$ & -22.92 & 0.62 & 0.70 & 0.19 & 0.017 \\
2.5--3.5 & $-22.97^{-0.27}_{+0.48}$ & $10.64^{+0.14}_{-0.25}$ & $ 8.88^{+0.04}_{-0.08}$ & -23.39 & 0.77 & 0.87 & 0.16 & 0.012 \\
\enddata
\label{tab:cutvalues}
\tablenotetext{a}{Redshift range}
\tablenotetext{b}{Inferred AB absolute magnitude threshold and 15--85\% confidence uncertainty interval, measured from the data.  Confidence interval does not include uncertainty due to cosmic variance in the field survey.}
\tablenotetext{c}{Stellar mass corresponding to column (b).}
\tablenotetext{d}{Oxygen abundance corresponding to column (c), converted using the analytic mass-metallicity relation of \cite{Zahid+2014}.}
\tablenotetext{e}{Absolute magnitude threshold predicted, assuming a metallicity cutoff of 12+log[O/H] = 8.94 at all redshifts.}
\tablenotetext{f}{Fraction of cosmic star-formation at that epoch which readily produces GRBs.}
\tablenotetext{g}{$p$ value calculated from a two-sided K-S test betwen the GRB host magnitude distribution and the SFR-weighted galaxy distribution clipped at the best-fit magnitude threshold.}
\tablenotetext{h}{$p$ value if clipped at the magnitude threshold corresponding to 12+log[O/H] = 8.94.}
\tablenotetext{i}{$p$ value if not clipped at all (if GRBs are assumed to be ``unbiased'' tracers of SFR).}
\end{deluxetable*}

Given that the original Zahid et al. relation was derived from a fit using only $z<1.7$ data the extension of their analytic prescription to higher redshifts is purely an extrapolation, and it is somewhat surprising that we continue to see agreement between our measured threshold and their mass-metallicity-redshift relation even at $z=2.25$ and (to some extent) at $z=3$.   The actual mass-metallicity relation at higher redshifts has been investigated directly by other surveys \citepeg{Savaglio+2005,Erb+2006,Maiolino+2008}\footnote{It has also been studied using GRB hosts themselves, comparing line-of-sight absorption metallicities to \emph{Spitzer} stellar masses \citep{Laskar+2011}.} but remains uncertain, in part because the ionization properties of high-redshift galaxies appear to be quite different than those at lower redshifts \citepeg{Steidel+2014,Sanders+2015}, putting the calibration of popular diagnostics and prescriptions developed at $z\sim0$ into doubt at these epochs.   Our observations may actually help shed some light on this topic:  given the good consistency between the GRB host luminosity function with the evolving mass-metallicity relation from $z\sim0.5$ to $z\sim2.0$ with a fixed, approximately Solar metallicity limit, the continued avoidance of GRBs from massive ($\sim10^{11} M_\odot$) galaxies out to $z\sim3$ suggests that massive systems are metal-enriched at or above at least the Solar value (and would favor the use of abundance diagnostics/calibrations consistent with this evaluation).  If so, this would be an interesting reversal:  while the metallicity sensitivity of the GRB progenitor has long been seen as an obstacle for its use to measure the cosmic star-formation rate density, this property may in the end turn out to be valuable by making GRBs a independent tracer of \emph{chemical} evolution.

Turning to low redshifts, although our inferred cut value provides a good explanation for the luminosity distrubution of galaxies over the full range of redshifts well-probed by our survey, it is somewhat higher than the effective metallicity cuts measured by spectroscopic surveys at low redshift.  For example, \citealt{Modjaz+2008} infer a cutoff value of $\sim$8.5 (on the \citealt{Kewley+2002} abundance scale, equivalent to 8.67 on KK04) based on four objects at $z\sim0.1$; the updated analysis of \cite{Graham+2013} finds reasonable agreement with this cut value for the bulk of their population (although fully 5 of their 14 hosts have measured metallicities above this.)  \cite{Wolf+2007} find a cutoff value of $8.7\pm0.3$ (KK04) using a photometric analysis of published low-$z$ hosts.  Given the large uncertainties on these earlier values (associated with the small sample size) they might not be inconsistent with our new measurement.   In addition, it must be remembered that the sample of local, low-luminosity GRBs has properties quite different from the cosmological population observed at high redshift (e.g., $>$5 orders of magnitude in $E_{\rm iso}$ and in observed volumetric rate; low-luminosity GRBs are also quasi-spherical and probably not jetted; \citealt{Bromberg+2011}).  
Most low-redshift samples are also still based primarily on pre-Swift GRBs whose selection is particularly heterogeneous and difficult to model; future studies of low-$z$ events from the more uniform, better-understood Swift population should help better constrain the true nature of the low-redshift population.

% Figure 7
\begin{figure*}
\centerline{
\includegraphics[scale=0.74,angle=0]{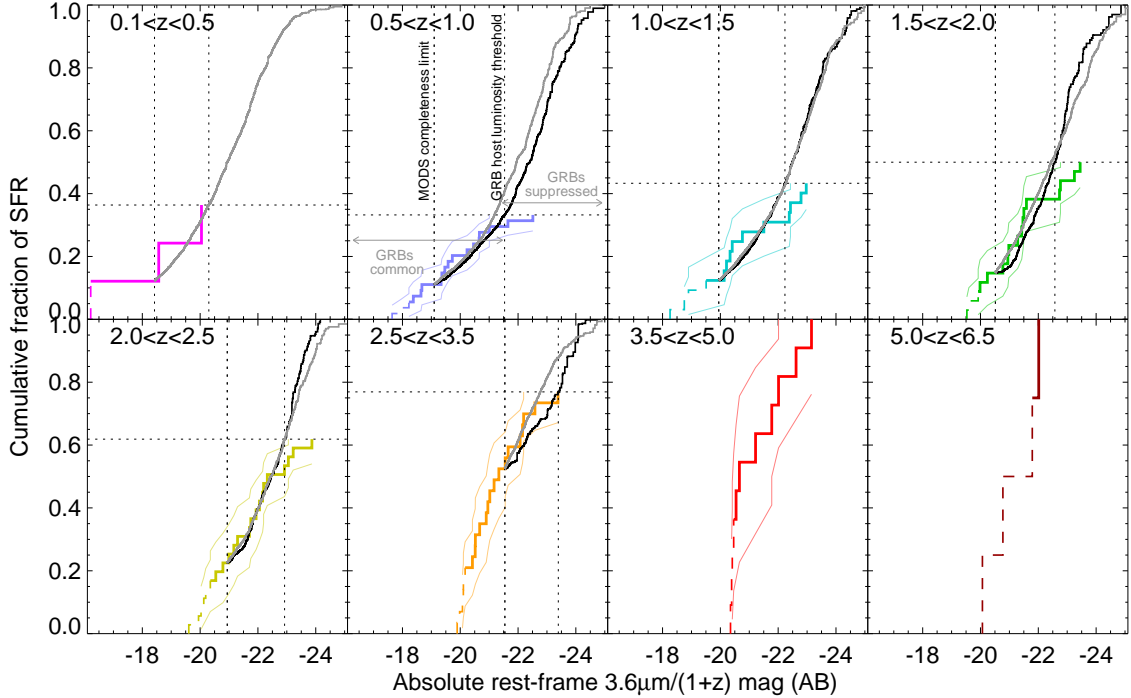}} 
\caption{Similar to Figure \ref{fig:lumdistvssfr}, but under a model in which GRB production is stifled above a certain luminosity threshold.  The threshold is determined via our procedure in \S \ref{sec:threshold} by marginalization using a K-S analysis, smoothed in redshift via a polynomial fit: this value is shown as the right vertical dotted line in each panel.  The threshold is also translated to the fraction of star-formation traced well by GRBs, indicated by the horizontal dashed line.  GRBs are poor tracers of star-formation at low redshifts (nearly all GRBs at $z < 1.5$ were produced by galaxies responsible for 50\% or less of cosmic star formation at that epoch).  At higher redshifts they are much better tracers, but are still not ``perfect'' (the $\sim$15\% of star-formation in the most luminous galaxies still produce few to no GRBs).}
\label{fig:lumdistvssfrcut}
\end{figure*}

% S3.5
\subsection{Peering Below the Surface:  The Fraction of Star Formation in IRAC-Undetected Galaxies}

If the GRB efficiency is close to uniform in galaxies with luminosities/metallicities below our inferred threshold, our observations can be used to provide an independent means of estimating what fraction of all star-formation occurs in the least luminous and lowest-mass galaxies at each redshift.  At the completeness level of the MODS survey (approximately $m_{3.6} = 24.25$, slightly shallower than the level achieved our host galaxy survey) the star-forming fraction in fainter galaxies can be read directly from our plots in Figure \ref{fig:lumdistvssfrcut} as the fraction where the completeness magnitude intersects the GRB host curve.  Given the depth of the survey, this fraction is unsurprisingly quite low at $z<2$: 10--20\% in every well-sampled redshift bin.   At higher redshifts the implied survey-detectability fraction drops significantly: the fraction of faint hosts is $\sim$30\% at $z=2.25$ and $\sim$50\% at $z=3$.  Our survey goes deeper than MODS at $z\sim3$ ($m_{\rm lim}\sim25$) and gives an even tighter constraint to this depth; only about 30\% of star-formation is in galaxies fainter than that level (about 3$\times$10$^9 M_\odot$).

Extending this constraint to even higher redshift ($z>3.5$) is contingent on the assumption that the GRB host distribution probes the entire galaxy population (i.e., that even the most massive galaxies have metallicities below our inferred threshold), an assessment we cannot test directly because no mass-complete field-galaxy samples exist at these redshifts.\footnote{Two other manuscripts were recently submitted on this topic, using the rest-frame UV luminosity distribution.  They come to different conclusions:  \citealt{Schulze+2015} show evidence for a GRB host distribution that may be skewed towards fainter hosts at $z\sim4$ (but not at $z\sim2-3$) compared to a uniform tracer, while \citealt{Greiner+2015} argue that the GRB host luminosity distribution follows the expected distribution.}   However, at the depths achieved on our deep high-$z$ fields ($m_{3.6} \sim 25.5$ mag, corresponding to $\sim$few $\times10^9 M_\odot$) we continue to detect approximately half the GRB population out to the redshift limit of the sample ($z\sim5.5$), indicating that about half of star-formation occurs in galaxies less luminous/massive than this threshold.  If GRBs continue to avoid high-mass galaxies even in this redshift range, the implied fraction of undetectable star-formation would be less than this.

% S3.6
\subsection{Correcting the GRB-Inferred Cosmic Star-Formation Rate History}

% Figure 8
\begin{figure*}
\centerline{
\includegraphics[scale=0.74,angle=0]{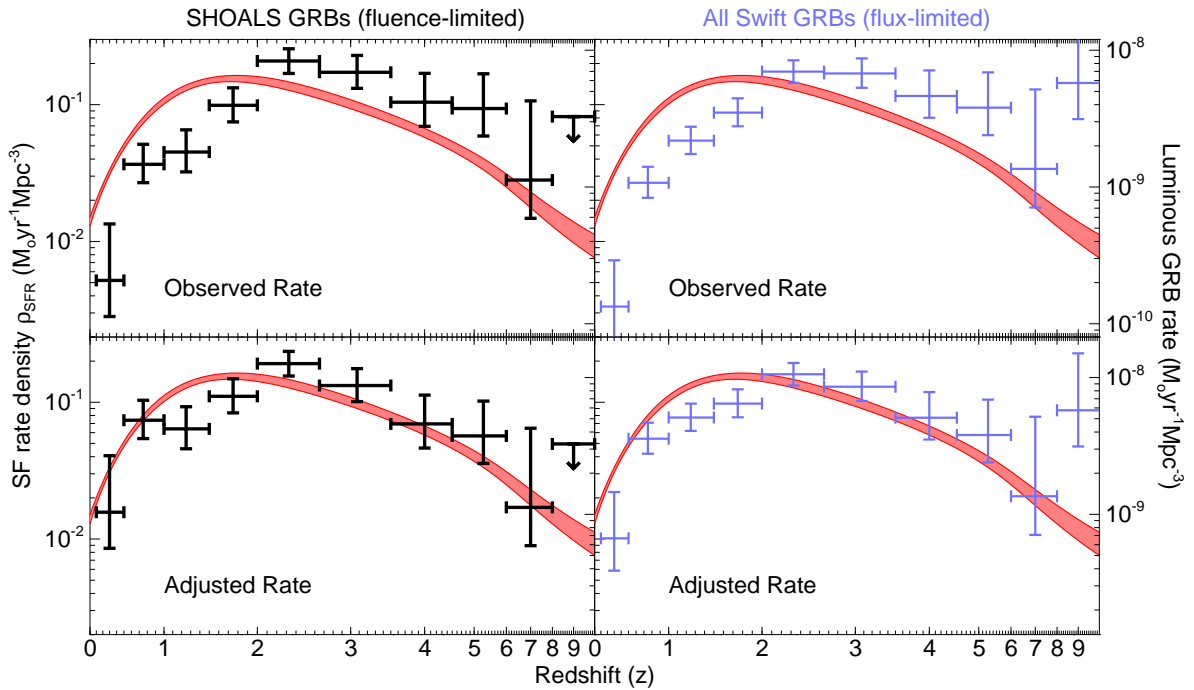}} 
\caption{The cosmic rate density of GRBs compared to that of star-formation.  The top panels show the raw, uncorrected GRB rate: the plot on the left is based on the SHOALS sample and uses the same data and calculation procedure as Paper I (based on $S$ and $E_{\rm iso}$); the plot on the right uses all \Swift GRBs from 2005--2014 (a larger but potentially biased sample) and uses the more traditional $f_{\rm av}$ and $L_{\rm iso, av}$.  Error bars show the 10--90\% binomial confidence interval \citep{Cameron2011}.  An ``excess'' in the GRB rate relative to the galaxy-inferred star-formation rate density \citep{Madau+2014,Robertson+2015} is seen at high-$z$.   In the bottom panels we have corrected these rates based on the observed tendency for GRBs to avoid luminous galaxies at low-$z$. Applying this correction produces good consistency in the respective rates, with the exception of the $z=8-10$ flux-limited bin.}
\label{fig:compareratez}
\end{figure*}

Figure \ref{fig:lumdistvssfrcut} also allows us to measure directly the fraction of star formation at each epoch that does vs. does not produce GRBs (except possibly at a much reduced rate)---this is simply the fraction of the plot below vs. above (respectively) the horizontal dotted line.  In the last column of Table \ref{tab:cutvalues} we provide this measurement, which is $\sim0.85$ at $z\sim3$ (indicating that GRBs track star-formation well in all galaxies except for those responsible for 15\% of cosmic star-formation) but drops to 0.5 at $z\sim1.5$ (indicating that half or more of the star-formation in the Universe does not produce GRBs, except perhaps in unusual circumstances) and even lower at lower redshifts.

This allows us to correct the cosmic GRB rate estimated in Paper I to examine whether or not the same model that explains the evolution of the \emph{luminosity} distribution can also explain the \emph{redshift} distribution and in particular the apparent excess of GRBs at high-$z$ (or equivalently, the deficiency at low-$z$).  We use a polynomial fit to interpolate between the redshift bins for the values in Table \ref{tab:cutvalues} and simply divide each bin by the resulting value of $f_{\rm SF}$.  

The result is plotted in Figure \ref{fig:compareratez} (left panels).  The correction procedure greatly reduces the discrepancy; while the $z>2$ data points are still above the SFRD curves the excess is less than a factor of two and always less than 2$\sigma$ significance.  

Since our GRB rate derivation procedure (using $E_{\rm iso}$ from a fluence-limited sample) differs from previous work, in the right panels of Figure \ref{fig:compareratez} we also show the rate derived in the more conventional fashion using the average flux/luminosity of each burst, $F = S / T_{\rm 90}$ (we continue use a $k$-correction to the 45-450 keV band instead of bolometric corrections).  Results are nearly identical, with the exception that this procedure recovers two $z>8$ bursts that were dropped by our fluence cut (090423 and 090429B; \citealt{Salvaterra+2009,Tanvir+2009,Cucchiara+2011}), which would imply a very high rate at this epoch.  This may imply different behavior entering the reionization era (in the burst rate or perhaps in burst properties---both events were remarkably short after taking into account time dilation) but may also be a fluke.  More searches for very high-$z$ GRBs will be needed to investigate this.

Otherwise, our analysis provides further support to the notion that the entire GRB rate behavior---both as a function of galaxy mass and as a function of redshift---can be explained in terms of the chemical history of the universe.  Above $z\gtrsim3$ GRBs track star-formation; below this point massive galaxies build up metals and cease GRB production, shifting the host luminosity distribution to fainter galaxies and reducing the cosmic GRB rate.

% S4
\section{Discussion and Conclusions}
\label{sec:conclusions}

In this paper we have demonstrated the signatures of cosmic evolution in the physical properties of the GRB host population across a vast redshift range from $z\sim0.3$ out to $z\sim6$, only 1 Gyr from the Big Bang.   The median host NIR luminosity does not evolve much between $z\sim5$ and $z\sim1.5$, but at lower redshifts ($z<1.5$) the average luminosity drops by over a factor of 10.   The $z<1.5$ host population is dramatically fainter than would be expected for an population that traces star-formation uniformly, and our observations require that the GRB rate per unit star-formation is greatly reduced in massive galaxies compared to low-mass galaxies.  At higher redshifts of $1.5<z<3.5$, the population is also slightly underluminous compared to a fully uniform tracer, but the difference is much more modest.  Consistent conclusions have been reached in recent years based on optical (rest-frame UV) data \citepeg{Tanvir+2012,Greiner+2015}; our study links these efforts using a single methodology across nearly all of cosmic history (see also \citealt{Schulze+2015}).

We emphasize that our target population was selected in an unbiased manner and our redshift completeness is very high (above 90\%), so our conclusions involve no systematic error associated with sample selection and minimal systematic error associated with redshift measurement.   In principle, since the galaxies that have escaped redshift measurement are all quite faint, or since we could in principle have misidentified a bright foreground system with the host galaxy in a small number of cases, the host galaxy population might actually be slightly \emph{fainter} than we infer.  But even this would leave our statistical comparion against star-forming galaxies effectively unaltered: the unknown-$z$ hosts are fainter than the MODS galaxies against which our comparison is based (and so would not enter into the comparison); likewise, any hypothetical faint background hosts (which would be rare in any case, not representing more than a few percent of the sample; see e.g. \citealt{Bloom+2002} or \citealt{Perley+2013a}) would have to be undetected in our Spitzer/optical/HST imaging and would likewise be dropped from the comparison.  In any case, there is \emph{no} possibility that the host galaxy population is brighter than we have reported, which firmly rules out any model for which the GRB rate is independent of environment.

Our results illustrate the importance of an unbiased approach similar to our own when addressing GRB demographics at high redshift.  Had dark bursts been omitted from our survey (or no campaign had been undertaken to measure their redshifts) nearly all objects in galaxies with a mass of $10^{10} M_\odot$ or above would have been dropped, and the mass/metallicity threshold we would have inferred would have been correspondingly much lower than the near-Solar value we estimate from this study.  Or, had we drawn our sample from the general Swift pool without imposing the optimized observability cuts discussed in Paper I to produce a high initial redshift completeness, a large fraction of the GRBs in fainter hosts would not be able to have their redshifts measured, producing an apparent host population biased towards \emph{brighter} galaxies.
These types of biases have bedeviled previous work on this topic, leading to the confusion regarding the degree to which GRBs do or do not track the star-formation rate.  (See, e.g., \citealt{Savaglio+2012} or \citealt{Hunt+2014} for recent, contrary views of the GRB host mass distribution using samples that were not drawn in an unbiased manner.)
We hope that our unbiased selection, large sample, and uniform observational strategy will clear this point up unambiguously and pave the way towards effectively using GRBs to probe star-formation and galaxy evolution within the (large) regime in which they do seem to track predictions well, and perhaps even using their metallicity dependence as a tool for understanding chemical evolution.

We hypothesized that the behavior seen in our sample---a strong shift with decreasing redshift towards a lower-mass host population that deviates from the general population of star-forming galaxies more strongly---was the result of time-evolution in the mass-metallicity relation in a manner similar to what was previously predicted by, e.g., \cite{Kocevski+2009}.   Using a simple model in which the GRB efficiency is constant at low metallicity but falls sharply above a maximum-metallicity threshold, we found good agreement for a redshift-independent threshold metallicity of 12 + log[O/H] = 8.94 (on the KK04 abundance scale, equivalent to 8.64 on the PP04 scale)--close to the Solar value.  The same model provides a good explanation for the GRB rate history, largely eliminating the previously-reported GRB rate ``excess'' at high-$z$ and producing a consistent picture (within a factor of 2) of the cosmic star-formation history with the results of high-redshift field surveys.  A hint of a possible upturn at $z\sim8$ will require future investigation.

The metallicity threshold we infer is somewhat, but not dramatically, higher than what has been reported by previous GRB host surveys (and assumed by \citealt{Kocevski+2009}).  On the other hand, it is unambiguously much higher than the metallicity upper limits predicted by single-star theoretical models \citepeg{MacFadyen+1999,Hirschi+2005,Yoon+2005,Langer+2006}---likely indicating that the progenitor is a binary system of some sort (see also \citealt{Trenti+2015}).  Future models for the progenitor formation process will need to consider why the GRB production rate appears to exhibit a sharp metallicity threshold but does so at a value much higher than originally predicted.

We emphasize that the suppression of the GRB rate in galaxies above our luminosity cut need not be total: GRBs in metal-rich environments are known to exist \citepeg{Levesque+2010zrich,Graham+2013,Kruehler+2015,Hashimoto+2015}, and even if the metal avoidance was total we would expect chemical inhomogeneity to smear the threshold somewhat \citep{Niino+2011}.  Indeed, we observe a few hosts with luminosities above our best-fit threshold at each redshift range---but always fairly close to the threshold itself, and in limited numbers that suggest that the rate in this type of host must be lower by about an order of magnitude comped to hosts below the threshold.

If our interpretation that the GRB efficiency is nearly constant below this threshold is correct, our results can also be used to constrain the importance of low-mass galaxies to cosmic star-formation.   We detect a large majority of the targeted hosts at low to moderate redshift (90\% at $z\sim1$ and 70\% at $z\sim2-3$), consistent with the notion that most star-formation occurs in galaxies in the range of masses detectable to \emph{Spitzer} at these distances (approximately $z>10^9 M_\odot$)---not in extremely low-mass dwarfs or outside galaxies, consistent with the present understanding of star and galaxy formation (but in contradiction to a recent study of the unresolved optical background; \citealt{Zemcov+2014}).  We continue to detect about half of the targets in the sample even out to $z\sim5.5$, indicating that we have not yet reached the epoch of cosmic history where most of cosmic star-formation shifts to small galaxies that are still being formed.  Deep rest-frame UV observations of the (not-unbiased) sample of known GRB hosts at $z>5.5$ have not yet detected any targets down to very faint levels \citep{Tanvir+2012}, which may suggest that the transition to low-mass galaxies occurs close to this epoch.

While our observations support the leading paradigm in which metallicity is the primary factor controlling the rate of GRB production as a fraction of star formation, they do not yet establish that it is the \emph{sole} driver.  The consistency of the host mass population with a metallicity-threshold model at many different redshifts and the similar consistency with the cosmic star-formation rate both suggest that other factors cannot be predominant unless they correlate with metallicity in very specific ways to mimic a metallicity effect.  However, other influences remain possible.
We \citepeg{Perley+2013a,Perley+2015a} have previously noted that even the intermediate-mass host population at $z\sim1-2$ shows other peculiarities compared to what are thought to be typical star-forming galaxies: their colors are bluer, their specific star-formation rates are higher, and they are more likely to be detected at radio wavelengths.   \cite{Kelly+2014} has noted that GRB hosts at $z\sim0.5$ appear to be unusually compact compared to supernova host galaxies of similar mass. 
Finally, the cutoff value we infer from our sample at $z=0.5-2.0$ is slightly in excess of what has previously been inferred from lower-$z$ hosts; this could simply reflect the limited size and selection of low-redshift samples but could also indicate the influence of a second parameter (sSFR, IMF, etc.)   
We are currently collecting a large library of multi-band optical, NIR, and radio observations to examine these possibilities in detail and will report the results in future papers associated with the survey.

\acknowledgments

This work is based on observations made with the  {\it Spitzer Space Telescope}, which is operated by the Jet Propulsion Laboratory, California Institute of Technology, under a contract with NASA.  Support for this work was provided by NASA through an award issued by JPL/Caltech associated with program GO-90062.  

Additional support for this work was provided by NASA through Hubble Fellowship grant HST-HF-51296.01-A awarded by the Space Telescope Science Institute, which is operated by the Association of Universities for Research in Astronomy, Inc., for NASA, under contract NAS 5-26555.  D.A.P. further acknowledges support from a Marie Sklodowska-Curie Individual Fellowship within the Horizon 2020 European Union (EU) Framework Programme for Research and Innovation (H2020-MSCA-IF-2014-660113).

The Dark Cosmology Centre is funded by the DNRF. The research leading to these results has received funding from the European Research Council under the European Union's Seventh Framework Program (FP7/2007-2013)/ERC Grant agreement no. EGGS-278202.  S.~Schulze acknowledges support from CONICYT-Chile FONDECYT 3140534, Basal-CATA PFB-06/2007, and Project IC120009 ``Millennium Institute of Astrophysics (MAS)´´ of Iniciativa Cient\'{\i}fica Milenio del Ministerio de Econom\'{\i}a, Fomento y Turismo.

We thank the referee for many helpful suggestions that significantly improved this manuscript.   We thank R.~Ellis and D.~A.~Kann for helpful comments.  We also wish to thank our many SHOALS collaborators involved in all aspects of the survey, and the entire \emph{Swift} team for making this project possible.

\clearpage
\LongTables

% Table 2
\begin{deluxetable*}{lllllrrrr}  %*
\tabletypesize{\small}
\tablecaption{Spitzer 3.6$\mu$m host galaxy photometry}
\tablecolumns{7}
\tablehead{
\colhead{GRB} &
\colhead{$z$} & 
\colhead{RA\tablenotemark{a}} & 
\colhead{dec\tablenotemark{a}} &
\colhead{$m_{\rm 3.6}$\tablenotemark{b}} &
\colhead{$M_{\rm 3.6/(1+z)}$\tablenotemark{c}} &
\colhead{log$_{10}$ $M_*$\tablenotemark{d}} &
\colhead{$t_{\rm exp}$\tablenotemark{e}} &
\colhead{Prog ID \tablenotemark{f}}
}
\startdata
050128  & $<$5.5                 & 14:38:17.71 & $-$34:45:54.6 &  $>$25.00          & $           $ &          &   5400 &  90062 \\
050315  & 1.9500                 & 20:25:54.16 & $-$42:36:02.1 &     23.43$\pm$0.08 & $     -21.29$ & $  9.77$ &   5400 &  90062 \\
050318  & 1.4436                 & 03:18:51.01 & $-$46:23:44.1 &  $>$25.22          & $    >-18.90$ & $< 8.63$ &   5400 &  90062 \\
050319  & 3.2425                 & 10:16:47.91 & $+$43:32:54.1 &     24.64$\pm$0.12 & $     -21.02$ & $  9.69$ &   7200 &  40599 \\
050401  & 2.8983                 & 16:31:28.79 & $+$02:11:14.0 &     24.29$\pm$0.39 & $     -20.89$ & $  9.61$ &   3700 &  40599 \\
050525A & 0.606                  & 18:32:32.58 & $+$26:20:22.4 &  $>$24.61          & $    >-17.64$ & $< 8.11$ &    900 &   3653 \\
050726  & $<$3.5                 & 13:20:11.95 & $-$32:03:51.2 &  $>$24.44          & $           $ &          &   5400 &  90062 \\
050730  & 3.9693                 & 14:08:17.10 & $-$03:46:17.6 &  $>$25.47          & $    >-20.54$ & $< 9.46$ &   7400 &  40599 \\
050802  & 1.7102                 & 14:37:05.84 & $+$27:47:12.3 &     24.86$\pm$0.29 & $     -19.60$ & $  9.00$ &   5400 &  90062 \\
050803  & 4.3$^{+0.6}_{-2.4}$    & 23:22:37.85 & $+$05:47:08.8 &     25.49$\pm$0.42 & $     -20.66$ & $  9.53$ &   5400 &  90062 \\
050814  & 5.3                    & 17:36:45.39 & $+$46:20:21.6 &     24.48$\pm$0.11 & $     -22.02$ & $ 10.22$ &   6900 &  40599 \\
050820A & 2.6147                 & 22:29:38.11 & $+$19:33:36.7 &     24.85$\pm$0.28 & $     -20.42$ & $  9.38$ &   3600 &  40599 \\
050822  & 1.434                  & 03:24:27.21 & $-$46:01:59.7 &     22.60$\pm$0.08 & $     -21.50$ & $  9.85$ &    540 &    272 \\
050904  & 6.295                  & 00:54:50.88 & $+$14:05:09.5 &  $>$24.98          & $    >-21.79$ & $<10.07$ &   7200 &  20000 \\
050922B & 4.9$^{+0.3}_{-0.6}$    & 00:23:13.30 & $-$05:36:17.8 &     24.59$\pm$0.22 & $     -21.78$ & $ 10.08$ &   5400 &  90062 \\
050922C & 2.1995                 & 21:09:33.08 & $-$08:45:30.2 &  $>$25.34          & $    >-19.61$ & $< 9.01$ &   3700 &  40599 \\
051001  & 2.4296                 & 23:23:48.77 & $-$31:31:24.0 &     23.20$\pm$0.08 & $     -21.94$ & $ 10.08$ &   5400 &  90062 \\
051006  & 1.059                  & 07:23:14.09 & $+$09:30:20.1 &     20.50$\pm$0.01 & $     -22.97$ & $ 10.58$ &   2700 &  90062 \\
060115  & 3.5328                 & 03:36:08.33 & $+$17:20:42.6 &     25.40$\pm$0.31 & $     -20.41$ & $  9.43$ &   7400 &  40599 \\
060202  & 0.785                  & 02:23:22.91 & $+$38:23:04.3 &     22.59$\pm$0.04 & $     -20.23$ & $  9.32$ &   5400 &  90062 \\
060204B & 2.3393                 & 14:07:14.93 & $+$27:40:36.3 &     22.74$\pm$0.05 & $     -22.32$ & $ 10.29$ &   5400 &  90062 \\
060210  & 3.9122                 & 03:50:57.37 & $+$27:01:34.0 &     23.37$\pm$0.06 & $     -22.62$ & $ 10.46$ &   7200 &  40599 \\
060218  & 0.0331                 & 03:21:39.69 & $+$16:52:01.8 &     20.77$\pm$0.04 & $     -15.01$ & $  7.20$ &    700 &  90062 \\
060306  & 1.559                  & 02:44:22.91 & $-$02:08:54.2 &     21.54$\pm$0.02 & $     -22.73$ & $ 10.50$ &   5400 &  90062 \\
060502A & 1.5026                 & 16:03:42.63 & $+$66:36:02.6 &  $>$24.67          & $    >-19.53$ & $< 8.96$ &   5400 &  90062 \\
060510B & 4.9                    & 15:56:29.48 & $+$78:34:12.1 &     25.15$\pm$0.17 & $     -21.22$ & $  9.81$ &  13000 &  20000 \\
060522  & 5.11                   & 21:31:44.86 & $+$02:53:09.7 &  $>$26.37          & $    >-20.07$ & $< 9.31$ &  13800 &  20000 \\
060526  & 3.2213                 & 15:31:18.34 & $+$00:17:04.9 &     25.48$\pm$0.29 & $     -20.17$ & $  9.30$ &   7200 &  40599 \\
060607A & 3.0749                 & 21:58:50.4  & $-$22:29:47.2 &  $>$25.05          & $    >-20.52$ & $< 9.45$ &   7200 &  40599 \\
060707  & 3.4240                 & 23:48:19.06 & $-$17:54:17.7 &     24.10$\pm$0.13 & $     -21.66$ & $  9.99$ &   6180 &  40599 \\
060714  & 2.7108                 & 15:11:26.41 & $-$06:33:58.2 &     25.23$\pm$0.26 & $     -20.11$ & $  9.25$ &   3600 &  40599 \\
060719  & 1.5320                 & 01:13:43.71 & $-$48:22:51.2 &     22.77$\pm$0.06 & $     -21.47$ & $  9.84$ &   1600 &  70036 \\
060729  & 0.5428                 & 06:21:31.78 & $-$62:22:12.0 &     24.06$\pm$0.19 & $     -17.95$ & $  8.31$ &   1600 &  90062 \\
060814  & 1.9229                 & 14:45:21.32 & $+$20:35:10.5 &     21.44$\pm$0.03 & $     -23.25$ & $ 10.82$ &    900 &  70036 \\
060908  & 1.8836                 & 02:07:18.41 & $+$00:20:31.3 &     24.73$\pm$0.20 & $     -19.92$ & $  9.15$ &   3600 &  40599 \\
060912A & 0.937                  & 00:21:08.12 & $+$20:58:17.7 &     21.56$\pm$0.03 & $     -21.65$ & $  9.91$ &   2700 &  90062 \\
060927  & 5.467                  & 21:58:12.01 & $+$05:21:48.6 &  $>$25.77          & $    >-20.78$ & $< 9.63$ &   7500 &  40599 \\
061007  & 1.2622                 & 03:05:19.59 & $-$50:30:02.3 &     23.71$\pm$0.15 & $     -20.13$ & $  9.22$ &   5400 &  90062 \\
061021  & 0.3463                 & 09:40:36.13 & $-$21:57:04.9 &  $>$24.78          & $    >-16.22$ & $< 7.63$ &   1600 &  90062 \\
061110A & 0.7578                 & 22:25:09.84 & $-$02:15:31.4 &     24.37$\pm$0.29 & $     -18.38$ & $  8.41$ &   1600 &  90062 \\
061110B & 3.4344                 & 21:35:40.40 & $+$06:52:33.9 &  $>$25.24          & $    >-20.52$ & $< 9.47$ &   7400 &  40599 \\
061121  & 1.3145                 & 09:48:54.54 & $-$13:11:43.2 &     21.50$\pm$0.01 & $     -22.43$ & $ 10.31$ &   5400 &  90062 \\
061202  & 2.253                  & 07:02:05.68 & $-$74:41:55.3 &     23.24$\pm$0.06 & $     -21.75$ & $  9.99$ &   5400 &  90062 \\
061222A & 2.088                  & 23:53:03.40 & $+$46:31:58.8 &     24.06$\pm$0.08 & $     -20.79$ & $  9.55$ &  18500 &  30000 \\
070110  & 2.3521                 & 00:03:39.24 & $-$52:58:27.3 &  $>$25.13          & $    >-19.95$ & $< 9.16$ &   3600 &  40599 \\
070129  & 2.3384                 & 02:28:00.91 & $+$11:41:04.2 &     23.01$\pm$0.03 & $     -22.05$ & $ 10.15$ &   6200 &  40598 \\
070223  & 1.6295                 & 10:13:48.40 & $+$43:08:00.6 &     23.59$\pm$0.08 & $     -20.77$ & $  9.53$ &   5400 &  90062 \\
070306  & 1.4959                 & 09:52:23.31 & $+$10:28:55.4 &     21.40$\pm$0.03 & $     -22.79$ & $ 10.53$ &    900 &  70036 \\
070318  & 0.840                  & 03:13:56.77 & $-$42:56:46.2 &  $>$24.30          & $    >-18.67$ & $< 8.54$ &   1600 &  80153 \\
070328  & 2.0627                 & 04:20:27.65 & $-$34:04:00.5 &     23.12$\pm$0.06 & $     -21.71$ & $  9.97$ &   5400 &  90062 \\
070419B & 1.9588                 & 21:02:49.78 & $-$31:15:49.1 &     23.30$\pm$0.09 & $     -21.43$ & $  9.84$ &   5400 &  90062 \\
070508  & 0.82                   & 20:51:11.72 & $-$78:23:04.8 &     22.46$\pm$0.07 & $     -20.46$ & $  9.41$ &   1600 &  70036 \\
070521  & 2.0865                 & 16:10:38.68 & $+$30:15:22.8 &     21.93$\pm$0.04 & $     -22.92$ & $ 10.63$ &   1600 &  70036 \\
070621  & $<$5.5                 & 21:35:10.05 & $-$24:49:02.5 &  $>$24.97          & $           $ &          &   5400 &  90062 \\
070721B & 3.6298                 & 02:12:32.96 & $-$02:11:40.8 &  $>$25.47          & $    >-20.39$ & $< 9.42$ &  10800 &  80054 \\
070808  & $\sim$1.35$\pm$0.85    & 00:27:03.30 & $+$01:10:35.4 &     23.57$\pm$0.10 & $     -20.41$ & $  9.35$ &   5400 &  90062 \\
071020  & 2.1462                 & 07:58:39.78 & $+$32:51:39.6 &  $>$24.36          & $    >-20.54$ & $< 9.43$ &   3600 &  80054 \\
071021  & 2.4520                 & 22:42:34.32 & $+$23:43:06.2 &     22.10$\pm$0.05 & $     -23.05$ & $ 10.68$ &   1600 &  70036 \\
071025  & 4.8$^{+0.4}_{-0.4}$    & 23:40:17.07 & $+$31:46:42.6 &     24.32$\pm$0.15 & $     -22.01$ & $ 10.18$ &   4500 &  80153 \\
071112C & 0.8227                 & 02:36:50.97 & $+$28:22:16.5 &     23.59$\pm$0.11 & $     -19.34$ & $  8.89$ &   2700 &  90062 \\
080205  & 2.72$^{+0.24}_{-0.74}$ & 06:33:00.69 & $+$62:47:32.0 &     23.79$\pm$0.13 & $     -21.55$ & $  9.91$ &   5400 &  90062 \\
080207  & 2.0858                 & 13:50:02.97 & $+$07:30:07.3 &     20.98$\pm$0.02 & $     -23.87$ & $ 11.11$ &   1600 &  50562 \\
080210  & 2.6419                 & 16:45:04.01 & $+$13:49:35.6 &  $>$24.62          & $    >-20.67$ & $< 9.50$ &   1600 &  80153 \\
080310  & 2.4274                 & 14:40:13.80 & $-$00:10:30.7 &     23.84$\pm$0.25 & $     -21.29$ & $  9.78$ &   3500 &  80054 \\
080319A & 2.0265                 & 13:45:20.01 & $+$44:04:48.4 &     22.63$\pm$0.03 & $     -22.16$ & $ 10.21$ &   5400 &  90062 \\
080319B & 0.9382                 & 14:31:41.00 & $+$36:18:08.6 &     24.57$\pm$0.22 & $     -18.64$ & $  8.50$ &   3600 &  11116 \\
080325  & 1.78                   & 18:31:34.22 & $+$36:31:23.7 &     21.09$\pm$0.02 & $     -23.45$ & $ 10.91$ &   1600 &  70036 \\
080411  & 1.0301                 & 02:31:55.19 & $-$71:18:07.3 &  $>$24.66          & $    >-18.75$ & $< 8.54$ &   2700 &  90062 \\
080413A & 2.4330                 & 19:09:11.75 & $-$27:40:40.4 &  $>$24.15          & $    >-20.99$ & $< 9.64$ &   3600 &  80054 \\
080413B & 1.1014                 & 21:44:34.67 & $-$19:58:52.3 &     23.22$\pm$0.09 & $     -20.33$ & $  9.30$ &   2700 &  90062 \\
080430  & 0.767                  & 11:01:14.71 & $+$51:41:08.4 &     24.55$\pm$0.34 & $     -18.22$ & $  8.33$ &   1600 &  90062 \\
080603B & 2.6892                 & 11:46:07.67 & $+$68:03:39.8 &     25.43$\pm$0.24 & $     -19.89$ & $  9.15$ &  10800 &  11116 \\
080605  & 1.6403                 & 17:28:30.04 & $+$04:00:55.7 &     21.61$\pm$0.10 & $     -22.77$ & $ 10.53$ &   1600 &  80153 \\
080607  & 3.0368                 & 12:59:47.24 & $+$15:55:10.8 &     22.96$\pm$0.06 & $     -22.58$ & $ 10.45$ &   4500 &  70036 \\
080710  & 0.8454                 & 00:33:05.63 & $+$19:30:05.4 &  $>$23.52          & $    >-19.47$ & $< 8.95$ &   2700 &  90062 \\
080721  & 2.5914                 & 14:57:55.83 & $-$11:43:24.6 &  $>$24.30          & $    >-20.96$ & $< 9.63$ &   3600 &  80054 \\
080804  & 2.2045                 & 21:54:40.18 & $-$53:11:04.8 &     24.76$\pm$0.23 & $     -20.19$ & $  9.28$ &   3600 &  80054 \\
080805  & 1.5042                 & 20:56:53.44 & $-$62:26:39.4 &     22.70$\pm$0.08 & $     -21.50$ & $  9.86$ &   1600 &  80153 \\
080810  & 3.3604                 & 23:47:10.49 & $+$00:19:11.5 &     23.57$\pm$0.07 & $     -22.15$ & $ 10.24$ &  10800 &  80054 \\
080916A & 0.6887                 & 22:25:06.22 & $-$57:01:22.6 &     22.81$\pm$0.10 & $     -19.73$ & $  9.12$ &   1600 &  90062 \\
080928  & 1.6919                 & 06:20:16.83 & $-$55:11:58.7 &  $>$24.20          & $    >-20.24$ & $< 9.29$ &   1600 &  80153 \\
081008  & 1.967                  & 18:39:49.87 & $-$57:25:53.0 &  $>$24.75          & $    >-19.98$ & $< 9.18$ &   5400 &  90062 \\
081029  & 3.8479                 & 23:07:05.36 & $-$68:09:19.7 &  $>$25.61          & $    >-20.35$ & $< 9.39$ &  10800 &  80054 \\
081109A & 0.9787                 & 22:03:09.57 & $-$54:42:40.4 &     20.79$\pm$0.02 & $     -22.51$ & $ 10.35$ &    500 &  80153 \\
081118  & 2.58                   & 05:30:22.22 & $-$43:18:05.6 &     25.27$\pm$0.41 & $     -19.98$ & $  9.18$ &   3600 &  80054 \\
081121  & 2.512                  & 05:57:06.14 & $-$60:36:09.8 &     25.09$\pm$0.34 & $     -20.11$ & $  9.24$ &   3600 &  80054 \\
081128  & $<$3.4                 & 01:23:13.09 & $+$38:07:38.8 &     25.20$\pm$0.31 & $           $ &          &   5400 &  90062 \\
081210  & 2.0631                 & 04:41:56.17 & $-$11:15:26.8 &     22.63$\pm$0.04 & $     -22.20$ & $ 10.23$ &   5400 &  90062 \\
081221  & 2.26                   & 01:03:10.17 & $-$24:32:52.0 &     21.78$\pm$0.03 & $     -23.22$ & $ 10.78$ &   1600 &  70036 \\
081222  & 2.77                   & 01:30:57.57 & $-$34:05:42.5 &     24.48$\pm$0.21 & $     -20.90$ & $  9.61$ &   3600 &  80054 \\
090313  & 3.375                  & 13:13:36.21 & $+$08:05:49.8 &  $>$23.54          & $    >-22.19$ & $<10.26$ &   1600 &  80153 \\
090404  & 3.0$^{+0.8}_{-1.8}$    & 15:56:57.49 & $+$35:30:57.5 &     22.14$\pm$0.04 & $     -23.38$ & $ 10.85$ &   1600 &  70036 \\
090417B & 0.345                  & 13:58:46.61 & $+$47:01:04.5 &     20.94$\pm$0.02 & $     -20.05$ & $  9.44$ &    500 &  70036 \\
090418A & 1.608                  & 17:57:15.17 & $+$33:24:20.9 &     23.39$\pm$0.11 & $     -20.95$ & $  9.61$ &   5400 &  90062 \\
090424  & 0.544                  & 12:38:05.11 & $+$16:50:14.4 &     21.35$\pm$0.03 & $     -20.66$ & $  9.62$ &   1600 &  90062 \\
090516A & 4.109                  & 09:13:02.61 & $-$11:51:14.6 &     22.92$\pm$0.04 & $     -23.15$ & $ 10.71$ &  10800 &  80054 \\
090519  & 3.85                   & 09:29:07.00 & $+$00:10:48.9 &  $>$25.50          & $    >-20.46$ & $< 9.44$ &  10800 &  80054 \\
090530  & 1.266                  & 11:57:40.50 & $+$26:35:37.9 &  $>$24.34          & $    >-19.51$ & $< 8.93$ &   5400 &  90062 \\
090618  & 0.54                   & 19:35:58.48 & $+$78:21:24.7 &     22.63$\pm$0.07 & $     -19.36$ & $  9.04$ &   1600 &  90062 \\
090709A & 1.8$^{+0.5}_{-0.7}$    & 19:19:42.63 & $+$60:43:39.3 &     23.58$\pm$0.09 & $     -20.98$ & $  9.63$ &   4500 &  70036 \\
090715B & 3.00                   & 16:45:21.61 & $+$44:50:20.4 &     23.45$\pm$0.06 & $     -22.07$ & $ 10.18$ &  10800 &  80054 \\
090812  & 2.452                  & 23:32:48.56 & $-$10:36:17.2 &  $>$24.80          & $    >-20.35$ & $< 9.35$ &   3600 &  80054 \\
090814A & 0.696?                 & 15:58:26.39 & $+$25:37:53.2 &     22.98$\pm$0.09 & $     -19.58$ & $  9.05$ &   1600 &  90062 \\
090926B & 1.24                   & 03:05:13.94 & $-$39:00:22.5 &     21.42$\pm$0.02 & $     -22.38$ & $ 10.28$ &   5400 &  90062 \\
091018  & 0.971                  & 02:08:44.74 & $-$57:32:54.1 &     22.28$\pm$0.04 & $     -21.01$ & $  9.62$ &   2700 &  90062 \\
091029  & 2.752                  & 04:00:42.62 & $-$55:57:20.0 &  $>$24.03          & $    >-21.34$ & $< 9.81$ &   3600 &  80054 \\
091109A & 3.076                  & 20:37:01.81 & $-$44:09:30.1 &     25.49$\pm$0.35 & $     -20.08$ & $  9.25$ &  10800 &  80054 \\
091127  & 0.490                  & 02:26:19.87 & $-$18:57:08.6 &     23.21$\pm$0.12 & $     -18.57$ & $  8.66$ &   1600 &  90062 \\
091208B & 1.0633                 & 01:57:34.11 & $+$16:53:22.9 &  $>$25.22          & $    >-18.26$ & $< 8.28$ &   2700 &  90062 \\
100305A &                        & 11:13:28.07 & $+$42:24:14.3 &  $>$25.27          & $           $ &          &   5400 &  90062 \\
100615A & 1.398                  & 11:48:49.34 & $-$19:28:51.5 &     23.84$\pm$0.14 & $     -20.21$ & $  9.27$ &   4500 &  70036 \\
100621A & 0.542                  & 21:01:13.02 & $-$51:06:22.4 &     21.35$\pm$0.03 & $     -20.65$ & $  9.61$ &    900 &  70036 \\
100728B & 2.106                  & 02:56:13.46 & $+$00:16:52.1 &  $>$24.74          & $    >-20.13$ & $< 9.25$ &   3600 &  80054 \\
100802A & $<$3.1                 & 00:09:52.43 & $+$47:45:18.5 &     25.47$\pm$0.43 & $           $ &          &   5400 &  90062 \\
100814A & 1.44                   & 01:29:53.55 & $-$17:59:43.7 &     23.35$\pm$0.07 & $     -20.76$ & $  9.52$ &   5400 &  90062 \\
110205A & 2.22                   & 10:58:31.20 & $+$67:31:31.0 &     23.80$\pm$0.10 & $     -21.17$ & $  9.72$ &  11100 &  90062 \\
110709B & 2.09?                  & 10:58:37.11 & $-$23:27:16.8 &     24.82$\pm$0.31 & $     -20.03$ & $  9.20$ &   5400 &  90062 \\
120119A & 1.728                  & 08:00:06.95 & $-$09:04:53.6 &     22.89$\pm$0.07 & $     -21.59$ & $  9.91$ &   5400 &  90062 \\
120308A & $<$3.7                 & 14:36:20.12 & $+$79:41:11.9 &     24.98$\pm$0.27 & $           $ &          &   5400 &  90062 \\
\enddata
\label{tab:spitzerphot}
\tablenotetext{a}{Location of IRAC aperture center (J2000).  For targets with a detected host at any waveband this is the host galaxy centroid; for non-detected hosts the best afterglow position is given.}
\tablenotetext{b}{Measured IRAC aperture magnitude (3.6$\mu$m, AB) and 1$\sigma$ uncertainty.}
\tablenotetext{c}{Absolute AB host magnitude at a wavelength of $\lambda=$3.6$\mu$m/(1+$z$).  For targets with photometric redshifts we assume the best-fit redshift given in the table.}
\tablenotetext{d}{Logarithm of the stellar mass, as derived from the Spitzer-measured luminosity.}
\tablenotetext{e}{IRAC exposure time in seconds.}
\tablenotetext{f}{Spitzer Program ID number.  Programs are: 272--PI Le Floc'h, ``Probing the Dark Side of the Cosmic Evolution Using Dark Gamma-Ray Bursts''. 3653--PI Garnavich, ``Gamma-Ray Burst Physics in the Spitzer/Swift Era''.  20000 and 30000--PI Berger, ``Gotcha! Using Swift GRBs to Pinpoint the Highest Redshift Galaxies''.   40598--PI Jones, ``GRBs as Beacons of Star Formation at High Redshifts''.  40599--PI Chary, ``Unveiling the Galaxy Counterparts of Damped Lyman-alpha Absorbers using GRB-DLAs''.  50562--PI Levan, ``The nature of dark gamma-ray burst host galaxies''.  70036--PI Perley, ``The Host Galaxies of Dust-Obscured Gamma-Ray Bursts''. 80054--PI Berger, ``Probing the z$>$2 Mass-Metallicity Relationship with Gamma-Ray Bursts''. 80153--PI Perley, ``Understanding the Environmental Dependence of High-z Dust with GRB Hosts''. 90062--PI Perley, ``Spitzer Observations of GRB Hosts: A Legacy Approach''.}
\end{deluxetable*}

\clearpage

%\bibliography{ref}{}

\bibliographystyle{apj}

\end{document}